\documentclass[traditabstract]{aa} 
\usepackage{graphicx}
\usepackage{txfonts}

\begin{document}

\title{The Globular Cluster Systems of Abell 1185\thanks{Based on observations with the NASA/ESA {\it Hubble Space Telescope} obtained at the Space Telescope Science Institute (STScI), which is operated by the Association of Universities for Research in Astronomy, Inc., under NASA contract NAS 5-26555.}
}

\author{Michael J. West\inst{1}
          \and
          Andr\'es Jord\'an\inst{2}
          \and
          John P. Blakeslee\inst{3}
          \and
          Patrick C\^ot\'e\inst{3}
          \and
          Michael D. Gregg\inst{4,5}
          \and
          Marianne Takamiya\inst{6}
           \and
          Ronald O. Marzke\inst{7}
                    }

\institute{
           European Southern Observatory, Alonso de Cordova 3107, Vitacura, 
           Santiago, Chile\\
              \email{mwest@eso.org}
         \and
             Departamento de Astronom\'ia y Astrof\'isica, Pontificia Universidad Cat\'olica de Chile,
             Casilla 306, Santiago 22, Chile  \\
             \email{ajordan@astro.puc.cl}
          \and
              Herzberg Institute of Astrophysics, National Research Council of Canada, Victoria, 
              BC V9E 2E7, Canada \\
               \email{ john.blakeslee@nrc-cnrc.gc.ca, patrick.cote@nrc-cnrc.gc.ca}  
                \and
              Institute for Geophysics and Planetary Physics, Lawrence Livermore National 
              Laboratory, L-413, Livermore, CA 94550, USA \\
               \email{gregg@igpp.ucllnl.org}    
               \and
              Department of Physics, University of California, Davis, CA 956160, USA \\
              \email{gregg@igpp.ucllnl.org}                     
               \and
              Department of Physics and Astronomy, University of Hawaii, Hilo, HI 96720, USA \\
               \email{ mtakamiya@hawaii.edu}    
               \and
              Department of Physics and Astronomy, San Francisco State University, San Francisco, CA 94132-4163, USA \\
               \email{ marzke@stars.sfsu.edu}   
                }
\date{Received October 15, 2010}
 
\abstract{
We examine the properties of a previously discovered population of globular clusters in the heart of the rich galaxy cluster Abell 1185 that might be intergalactic in nature.   Deep images obtained with the {\em Advanced Camera for Surveys} ({\em ACS}) aboard {\em Hubble Space Telescope} ({\em HST}) confirm the presence of  
$\sim  1300$ globular clusters brighter than  $I_{F814W} \simeq 27.3$ mag in a field devoid of any large galaxies.  
The luminosities and colors of these objects are found to be similar to those of metal-poor globular clusters observed in many galaxies to date.  
Although a significant fraction of the detected globular clusters undoubtedly reside in the outer halos of galaxies adjacent to this field, detailed modeling of their distribution suggests that the majority of these objects are 
likely to be intergalactic, in the sense that they are not gravitationally bound to any individual galaxy. 
We conclude that the true nature and origin of the globular
cluster population in the core of A1185 $-$ galactic residents or intergalactic
wanderers $-$ remains uncertain, and suggest how future observation could resolve this ambiguity.}
   
\keywords{galaxies: elliptical and lenticular, cD --
                galaxies: formation --
                galaxies: interactions --
                galaxies: clusters: individual: Abell 1185 --
                globular clusters: general}
                 
\maketitle

\section{Introduction}

Globular clusters are ubiquitous in galaxies, found in all but the smallest dwarfs (see, e.g., Figure 3 of Peng et al. \cite{peng2008}). For this reason they have long played a key role in astronomers' quest to understand the origin and evolution of galaxies  (see West et al. \cite{west2004} and Brodie \& Strader \cite{bs2006} for reviews).  

However, a growing body of observational evidence also hints at the existence of a population of intergalactic globular clusters (hereafter IGCs) that reside outside of galaxies.
Half a century ago, van den Bergh (\cite{vdb1958}) speculated that ``about one third of all the globular clusters are of the intergalactic type'' and numerous studies since then have lent support to the idea that at least some globular clusters are not gravitationally bound to individual galaxies  (e.g., Muzzio et al. \cite{muzzio1984}; White \cite{white1987}; West et al. \cite{west1995}; Bassino et. al. \cite{bassino2003}; Hilker \cite{hilker2003}; Jord\'an et al. \cite{jordan2003}; Williams et al. \cite{williams2007}; Schuberth et al. \cite{schuberth2008}; Coenda, Muriel \& Donzelli \cite{coenda2009}; Gregg et al. \cite{gregg2009}; Lee et al. \cite{lee2010}; Peng et al. \cite{peng2011}).  

   \begin{figure*}
   \centering
  \includegraphics[width=17cm]{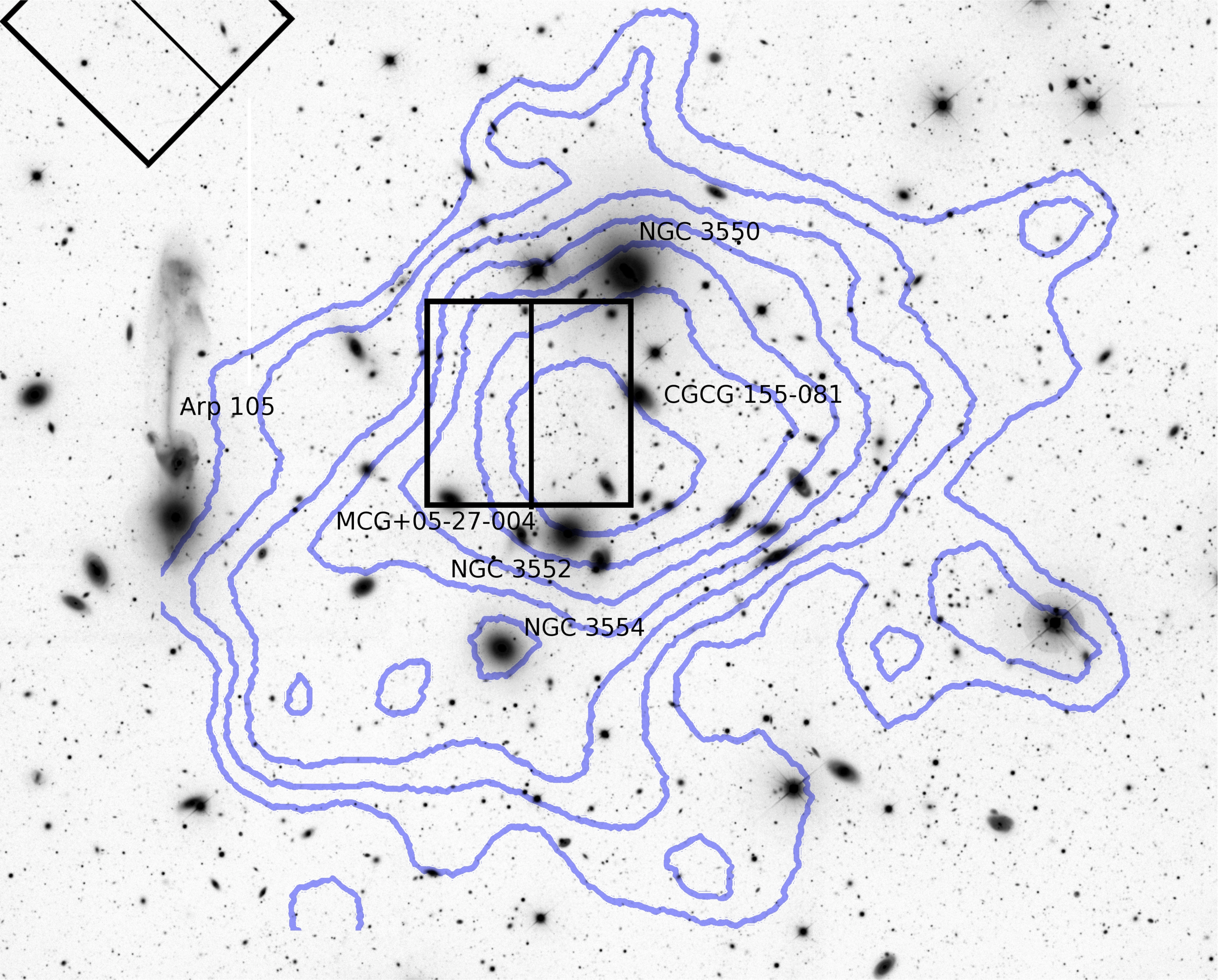}
   \caption{A composite $BVR$ image of the rich cluster A1185 obtained with the CFH12k camera on CHFT (see Andreon et al.  \cite{andreon2007}) showing the locations of our ACS fields.  The field of view is $\sim 20^{\prime} \times 16^{\prime}$, corresponding to $\sim 0.8$ Mpc $\times$ $0.64$ Mpc at the distance of A1185. North is up and East is to the left. Our  primary ACS field is located just below the brightest cluster galaxy, NGC 3550.  A second ACS field located $\sim 9^{\prime}$ ($\sim 350$ kpc) to the northeast was used as a control field, as discussed in the text.  Several of the brightest galaxies near the ACS field are identified, including the spectacular interacting galaxy pair Arp 105.  X-ray contours from Mahdavi et al. \cite{mahdavi1996} based on {\em Einstein} observations are overlaid on the image. 
 }
     \label{CFHT}
    \end{figure*}

Little is known at present about the nature or origin of IGCs.  One possibility is that they are normal globular clusters that were stripped from galaxies during gravitational interactions, the detritus of cosmic evolution.  
Globular clusters, being among the densest stellar systems, are likely to survive the partial or even complete disruption of their parent galaxy, resulting in an ever-growing population of orphaned star clusters that roam the space between galaxies.    Alternatively, some globular clusters might have been born in intergalactic space without ever residing in galaxies, formed perhaps from the gravitational instability of primordial density fluctuations (e.g., Peebles \& Dicke \cite{peebles1968}; Peebles \cite{peebles1984}; West \cite{west1993}; Yahagi \& Bekki \cite{yahagi2005}; Griffen et al. \cite{griffen2010} or thermal instabilities in cooling flows and other gas-rich environments  (e.g., Fabian et al. \cite{fabian1984}; Fall \& Rees \cite{fall1985}; Cen \cite{cen2001}; Griffen et al. \cite{griffen2010}).

Whatever their origin, 
IGCs are likely to be most abundant in the extreme environments of galaxy clusters,  where 
tidal forces and collisions between galaxies can pluck globular clusters from their halos
or the dense local conditions might have allowed the birth of massive star clusters from intergalactic gas.   Unfortunately, searches for IGCs in galaxy clusters are hindered by the fact that most have a giant elliptical at their center where the density of IGCs is expected to be greatest, making it difficult to distinguish genuine IGCs from the galaxy's intrinsic globular cluster population. 
  
An exception is Abell 1185 ($z = 0.033$), a rarity among low-redshift galaxy clusters because its brightest member galaxy (NGC 3550) is offset by 
$\sim 150$ kpc 
from the peak of the x-ray gas emission that presumably marks the centre of the cluster mass distribution (Mahdavi et al. 1996).   Hence, A1185 offers an ideal opportunity to search directly for IGCs in the core of a relatively nearby cluster of galaxies.  Tentative evidence of such a population was presented by Jord\'an et al. (2003), who detected 99 candidate IGCs in {\em HST/WFPC2} images of a field in the center of A1185.  However, their data suffered from incompleteness at magnitudes fainter than $I_{F814W} \simeq 25.5$ mag, allowing only the brightest $\sim 10\%$ of the expected globular cluster population to be detected at the distance of A1185.

In this paper, we analyze deep F555W and F814W {\it ACS} images of the core of A1185.  We confirm the presence of a substantial population of globular clusters in this field and extend the results of Jord\'an et al. (2003) by presenting the first observations of the luminosity function and color distribution of these objects.
The observations and data reductions are described in $\S 2$.  Results from our analysis are presented in $\S 3$, and their implications are discussed in $\S 4$.  We assume a standard  cosmology with $H_0$ = 72 km s$^{-1}$ Mpc$^{-1}$  and a distance modulus $m-M \simeq 35.7$ mag to A1185.  At this distance, an angle of $1\arcmin$ subtends a projected linear scale of 65 kpc, and the $0.05\arcsec$ pixel scale of the {\em ACS/WFC} corresponds to a physical scale of 32 pc.

   \begin{figure*}
   \centering
   \includegraphics[width=15cm]{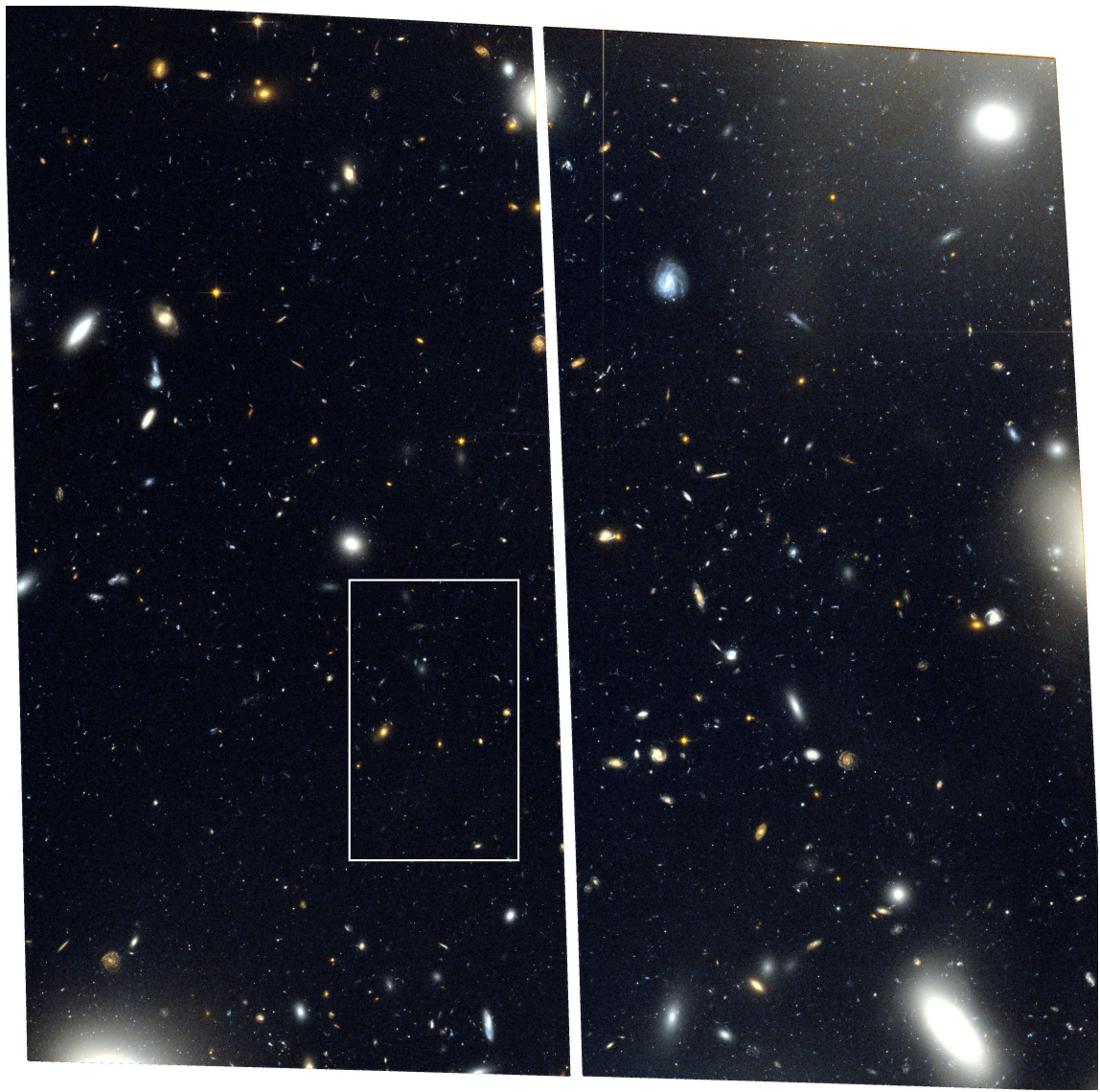}
   \caption{{\em ACS} image of the central field in Abell 1185, produced using the $V_{F555W}$ and $I_{F814W}$ images and averaging them to produce the third color.  The brightest cluster galaxy, NGC 3550, is located just off the upper right corner of the image.  The field of view is 
   $3.4\arcmin \times 3.4\arcmin$, 
   corresponding to a spatial scale of $\sim 220$ kpc on a side.  North is up and east is to the left.  The white box indicates the region shown in Fig. \ref{corefield}.}
     \label{a1185a996}
    \end{figure*}

\section{Observations and Data Reductions}
  
Deep images of a field near the location of the peak x-ray emission in A1185 (RA = 11h 10m 46s, DEC = $+28^{\circ} 43^{\prime} 57^{\prime\prime}$) were obtained using the {\em ACS} Wide Field Channel (WFC) onboard {\em HST} as part of program GO-9438, which was allocated 15 orbits of time (see Figures \ref{CFHT} and \ref{a1185a996}).  Two filters were used, F555W and F814W (roughly equivalent to Johnson $V$ and Cousins $I$), with total exposure times of 15,540 s and 23,160 s, respectively.   The longer integration time of the $I_{F814W}$ images was chosen to ensure 
 complete detection of the brightest half of the globular cluster luminosity function (GCLF) at the distance of A1185, which is assumed to have a Gaussian distribution with a peak at $M_V \simeq -7.4$ mag (e.g., Harris 2001; Jord\'an et al. 2007) and hence should occur at an apparent magnitude $I_{F814W} \simeq 27.3$ mag for typical globular cluster colors ($V-I \simeq 1$).  The relatively shallower limit of the $V_{F555W}$ images was intended to provide color information for brighter globular clusters only. With a scale of $0.049\arcsec$ per pixel, the images have a $202\arcsec \times 202\arcsec$ field of view corresponding to $\sim 130$ kpc on a  side at the distance of A1185 (for an assumed distance modulus $m-M = 35.7$).   Individual exposures were not dithered because the small gap between the {\em ACS} chips does not adversely affect our ability to detect globular clusters over most of the field.  
 
Images were processed using the Apsis ACS data pipeline (Blakeslee \cite{blakeslee2003}), which produces final sky-subtracted images in each filter after distortion correction, image alignment, cosmic ray removal and image combination using sub-pixel drizzling (Fruchter \& Hook {\cite{fruchter2002}) with a final pixel scale of $0.05^{\prime\prime}$.  A combined $V_{F555W}+I_{F814W}$ image is shown in Fig. \ref{a1185a996}.   Weight images were also constructed for each output science image; these give the total uncertainty for each pixel based on instrumental and photon shot noise (see Jord\'an et al. 2004).  These weight images provide a map of the {\it rms} variation in the final image that was used for object detection as described below.

\subsection{Object detection and classification}

Object detection was performed using the Source Extractor (SExtractor) software package (Bertin \& Arnouts \cite{bertin1996}) to produce a catalog of candidate globular clusters.  
At the distance of A1185, globular clusters are unresolved even with {\em ACS} and hence will appear as faint point sources (a typical globular cluster with half-light radius $r_h \simeq 3$ pc would have an angular size much less than one {\em ACS/WFC} pixel, however this is diffused over several pixels because of the {\em ACS} PSF which has a FWHM $\sim 2$ pixels). 

Objects were selected using a detection threshold of five contiguous pixels each at least $1.5 \sigma$ above background after convolving the image with a Gaussian filter ${\tt gauss_{-}2.0_{-}5 \times 5}$.
The {\it rms} maps produced by Apsis were used 
to weight each pixel based on its noise level by setting the ${\tt WEIGHT_{-}TYPE}$ parameter to ${\tt MAP_{-}RMS}$ when running SExtractor.   
The local background was estimated using a mesh size of $40 \times 40$ square pixels. 
These detection parameters were chosen after experimentation as providing the best compromise between sensitivity and reliability.    Object detection was done independently for the F555W and F814W images and the resulting object catalogs were then matched using a $0.1\arcsec$ matching radius.  

SExtractor assigns each detected object a classification parameter, CLASS, that ranges from $\sim 0$ for galaxies and other extended objects to $\sim 1$ for point sources. CLASS values are based on a neural network algorithm trained on simulated images (for details see Bertin \& Arnouts \cite{bertin1996}).  We consider point source candidates to have CLASS $\ge 0.8$.  Fig. \ref{class} shows that this criterion does an excellent job of discriminating globular clusters from galaxies and other non-point sources even at the faintest magnitudes.  Experiments in which artificial point sources of different magnitudes were added at random locations in the {\em ACS} field (see Sect. 2.3 for details) show that  fully $98\%$ of all input objects detected by SExtractor down to $I_{F814W} = 28$ mag were assigned CLASS $\ge$ 0.8.

   \begin{figure}
   \centering
   \includegraphics[width=9cm]{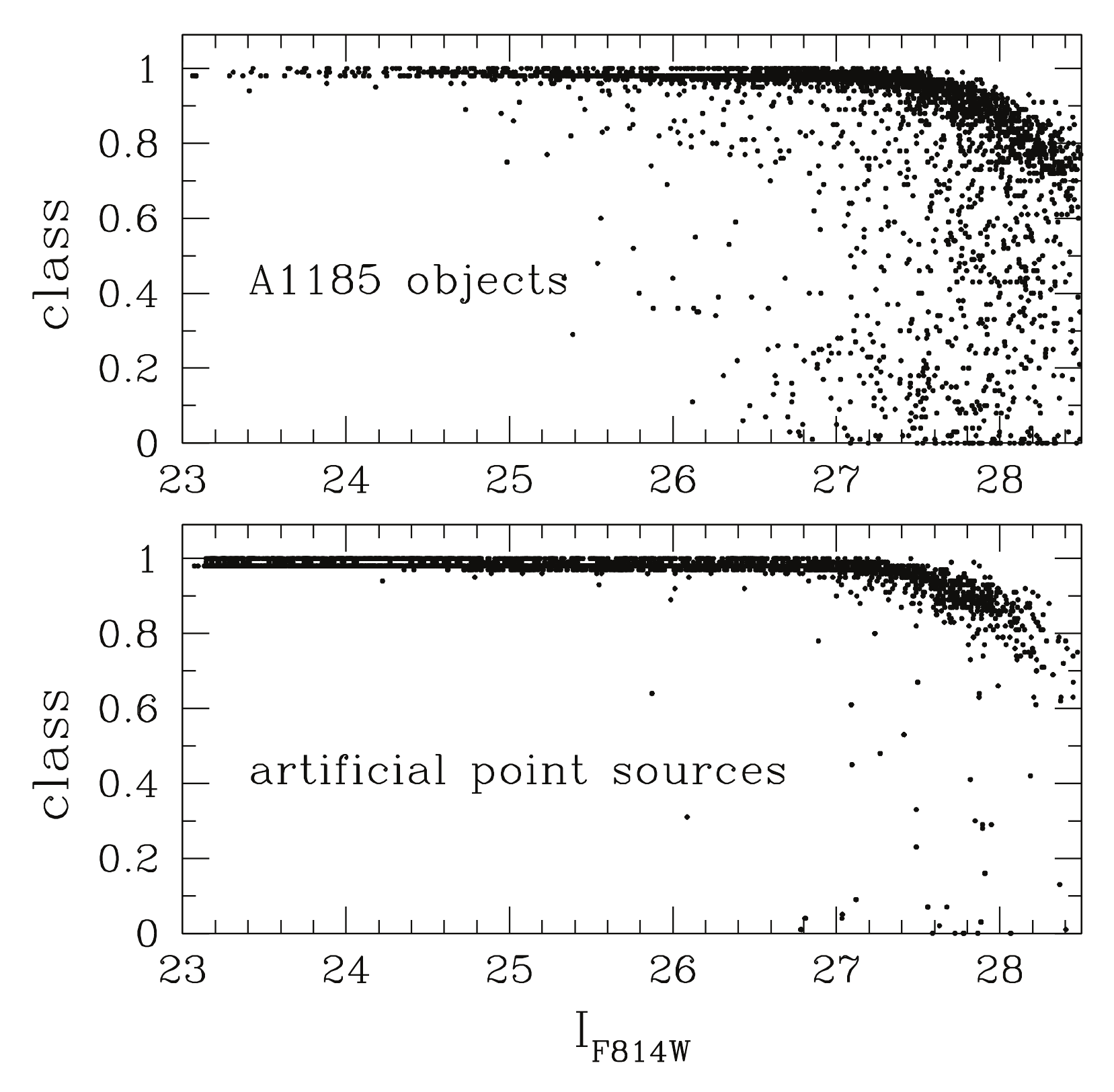}
      \caption{{\it Top:} SExtractor classes of all detected objects in the core of A1185. {\it Bottom:} SExtractor classes of   
      artificial point sources that were added to the field as described in Sect. 2.3.  Our adopted criterion of CLASS $\ge$ 0.8 for globular cluster candidates does an excellent job of 
      identifying point sources down to the faintest magnitudes.
              }
         \label{class}
   \end{figure}
%

\subsection{Photometric calibration}

Because globular clusters are unresolved at the distance of A1185, we used the ALLSTAR routine in the DAOPHOT software package (Stetson \cite{stetson1987}) to perform photometry of detected candidates based on PSF fitting, which provides more reliable magnitudes  than the default SExtractor magnitudes.   

To do this, the coordinates of SExtractor-detected objects were fed into DAOPHOT/ALLSTAR and instrumental magnitudes were then measured  at each location using PSF models determined empirically from stars on the final F555W and F814W images. 
We followed the procedures of Sirianni et al. (\cite{sirianni2005}) for aperture corrections and 
converted  instrumental magnitudes to the VEGAMAG system using the following zeropoints:
$$V_{555W} = -2.5~\mathrm{log}~f_{555W} + 25.724 \eqno(1a)$$
$$I_{814W} = -2.5~ \mathrm{log}~ f_{814W} + 25.501 \eqno(1b)$$
\noindent where $f_{555W}$ and $f_{814W}$ are the integrated fluxes in units of electrons per second in each filter.  
Reddening for the A1185 field was taken to be E(B - V) = 0.029 mag based on the DIRBE maps of Schlegel et al. (1998) and extinction ratios were taken from Sirianni et al. (\cite{sirianni2005}).  

\subsection{Completeness limits}

Object detection and classification become increasingly unreliable at faint magnitudes.  To assess our completeness limits, artificial stars of known magnitudes were added at random locations in the final {\em ACS} images and then the same SExtractor and DAOPHOT procedures described above were used to determine the fraction of stars recovered at each magnitude.    These experiments were performed by adding 100 stars at random to the {\em ACS} field using the 
DAOPHOT task ADDSTAR.  SExtractor was then run with the same detection parameters and the fraction of input stars that were recovered was recorded.
This process was repeated many times to generate  
the completeness function shown in Fig. \ref{completeness} for the F814W observations.  

These experiments show that object detection and classification are essentially $100\%$ complete down to the expected turnover magnitude of the GCLF at $I _{F814W} \simeq 27.3$ mag, and declines rapidly at fainter magnitudes.  
To avoid uncertainties caused by incompleteness at fainter magnitudes, we adopted a completeness limit at $I _{F814W} \simeq 27.3$ mag and consider only objects brighter than this in subsequent analysis.

   \begin{figure}
   \centering
   \includegraphics[width=9cm]{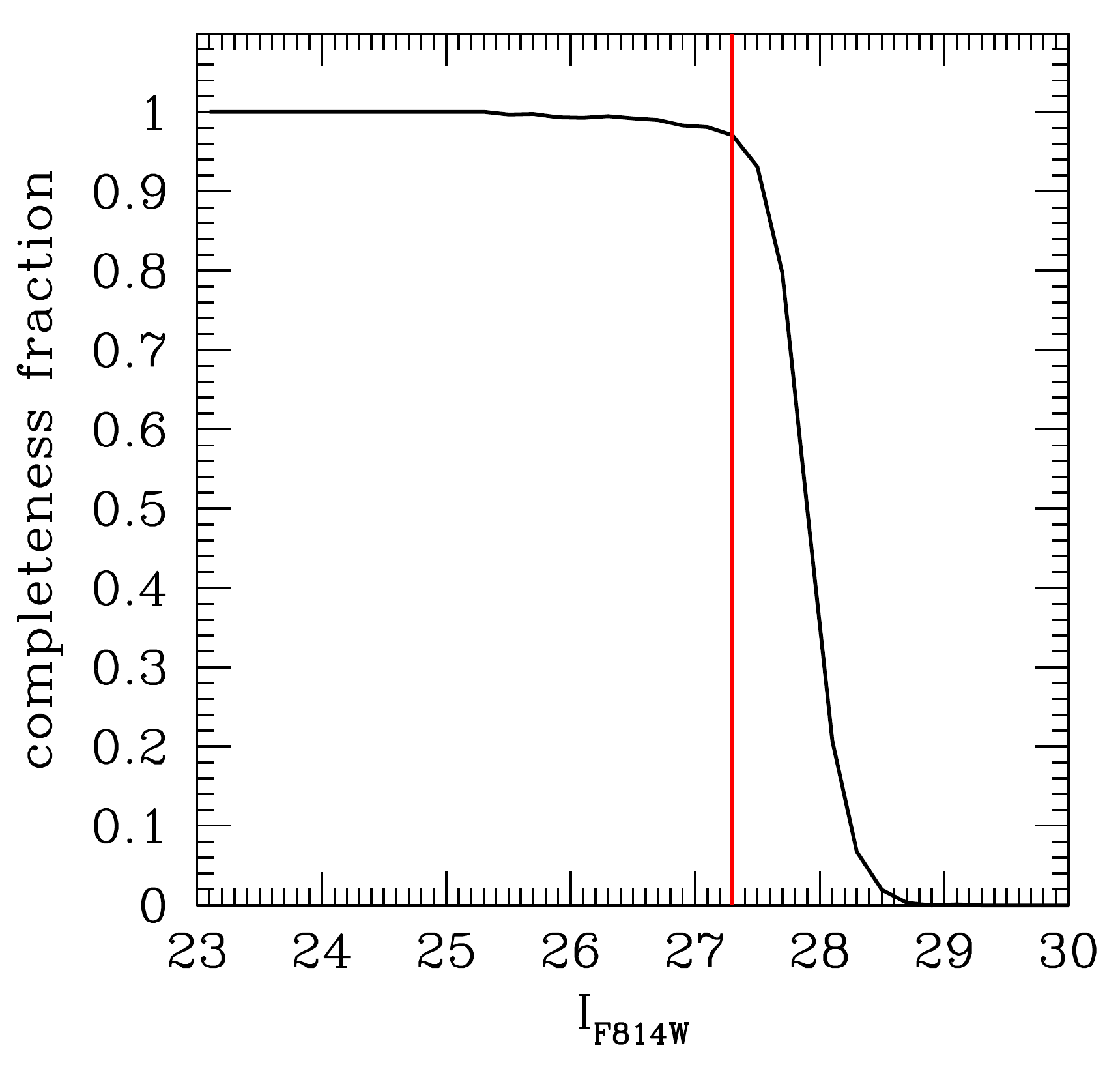}
      \caption{Completeness of point source detection as a function of apparent magnitude for the F814W image of the central region in A1185.  The red line shows the expected turnover magnitude of the globular cluster luminosity function at $I _{F814W} \simeq 27.3$ mag.
              }
         \label{completeness}
   \end{figure}
%

\section{Results}

Globular clusters detected in a small portion of the A1185 core field are shown in Fig. \ref{corefield}.
A total of 2285 such objects selected solely on the basis of their SExtractor classifications were found in the F814W image, of which 1422 are brighter than our adopted completeness limit $I_{F814W} = 27.3$ mag.  The projected spatial distribution of such objects is plotted in Fig. \ref{realxy2}.  Matched detections of 1124 objects were found in the  F555W image.  

It is clear from Fig. \ref{realxy2} that the two-dimensional distribution of IGC candidates is not uniform.  A prominent enhancement in the direction of NGC 3550 is easily seen, as are several 
other concentrations (e.g., around $x \simeq 500$, $y \simeq 300$ associated with the galaxy MCG+05-27-004 that lies just outside the ACS field).  
Hence we are undoubtedly seeing some contamination from globular clusters in the halos of galaxies, which we address below. 

   \begin{figure}
   \centering
   \includegraphics[width=9.0cm]{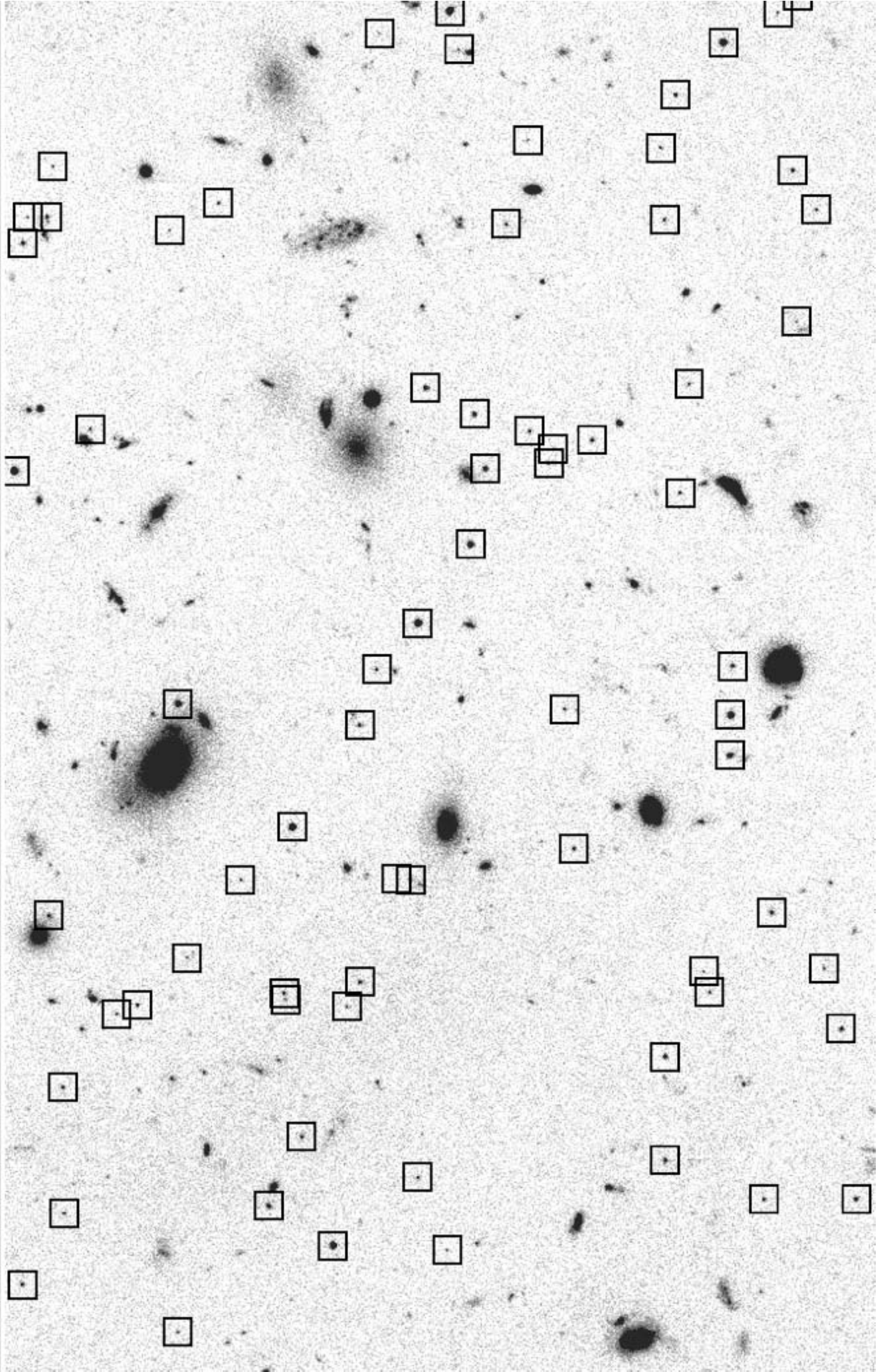}
   \caption{Intergalactic globular cluster candidates (in boxes) found in a small portion of the {\em ACS} field in the core of
      Abell 1185. The field of view is $\sim 0.3^{\prime} \times 0.5^{\prime}$.}
     \label{corefield}
    \end{figure}
    
   \begin{figure}
   \centering
   \includegraphics[width=9cm]{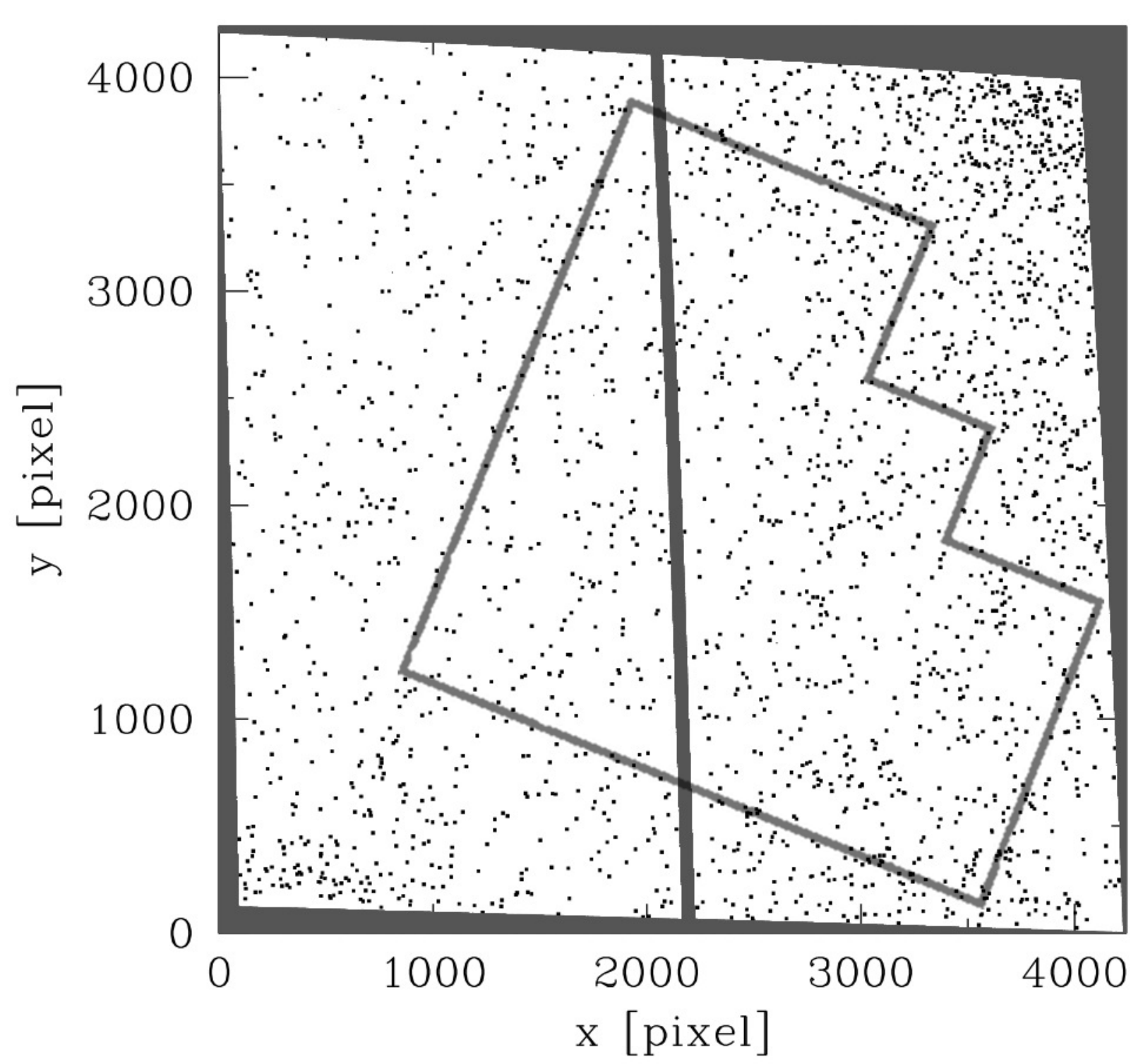} 
     \caption{The distribution of the 1422 intergalactic globular cluster candidates brighter than  $I_{F814W} = 27.3$ mag in the A1185 core field.  The field of view is 
   $3.4\arcmin \times 3.4\arcmin$, with north up and east to the left.  Enhancements 
     in the number of objects can be seen in a few regions, especially in the direction of the bright galaxy NGC 3550.  For comparison, the location of the smaller and shallower WFPC2 field studied by Jord\'an et al. 2003 is also shown. }
     \label{realxy2}%
    \end{figure}

\subsection{The number of IGCs in A1185}


Jord\'an et al. (2003) detected 99 IGC candidates in A1185 based on shallower {\it WFPC2} observations.  If intergalactic globular clusters have the same luminosity function as their counterparts in galaxies,  then scaling from the {\it WFPC2} results to the depth and area of our {\em ACS} field (assuming a constant density of objects  across the field) we would expect to find approximately 2000 to 4000 globular clusters in this region, with about half this number detected down to our adopted $100\%$ completeness limit at $I _{F814W} \simeq 27.3$ mag.  The detection of 1422 objects brighter than this magnitude limit is consistent with this estimate.

However, it is inevitable that some IGC candidates are actually foreground stars, unresolved background galaxies, and globular clusters in the halos of galaxies near the {\em ACS} field masquerading as IGCs.   Confirming 
the presence of IGCs in A1185 and estimating their number requires correcting for these various  sources of contamination.

To assess the contribution from stars and unresolved galaxies along the line of sight, four fields  observed with the same instrument, filter and comparable integration times were processed and analyzed in an identical manner as the A1185 core field.  The selected fields, which are all
at similar Galactic latitudes as the A1185 field, are devoid of bright galaxies and can therefore provide unbiased estimates of the number of contaminating point sources at different magnitudes. The four fields used were:

\begin{itemize}
\item A second field in A1185 (see Fig. \ref{CFHT}) was observed as part of our programme GO-10277.  This field is located $\sim 350$ kpc from the primary field and hence can provide a useful constraint on the spatial extent of the IGC distribution within A1185.  {\em ACS} observations with integration times of 17,816 seconds were obtained for both the F555W and F814W filters.  

\medskip
\item  F814W images of a field observed to study high-redshift Ly-$\alpha$ emitters (GO-9411) were extracted from the {\em HST} archive. 
This field is at Galactic latitude $b=68^{\circ}$, almost identical to that of A1185.  A subset of 16 images were combined to attain a total exposure time of 24,060 seconds.

\medskip
\item Two additional fields from the Gemini Deep Deep Survey (GO-9760) were obtained from the {\em HST} archive.  These are the SA02-1 and SA02-2 fields, which are located well away from the Galactic plane at $b= -61^{\circ}$.  The total integration time for both fields was 16,345 seconds with the F814W filter. 
\end{itemize}

Figure \ref{controlcompleteness} shows the completeness functions for each of the control fields determined from artificial star experiments.   All four show similar completeness limits, with nearly 100\% of objects detected down to $I_{F814W} \simeq 27.3$ mag, which is not surprising given their similar integration times.

   \begin{figure}
   \centering
   \includegraphics[width=9cm]{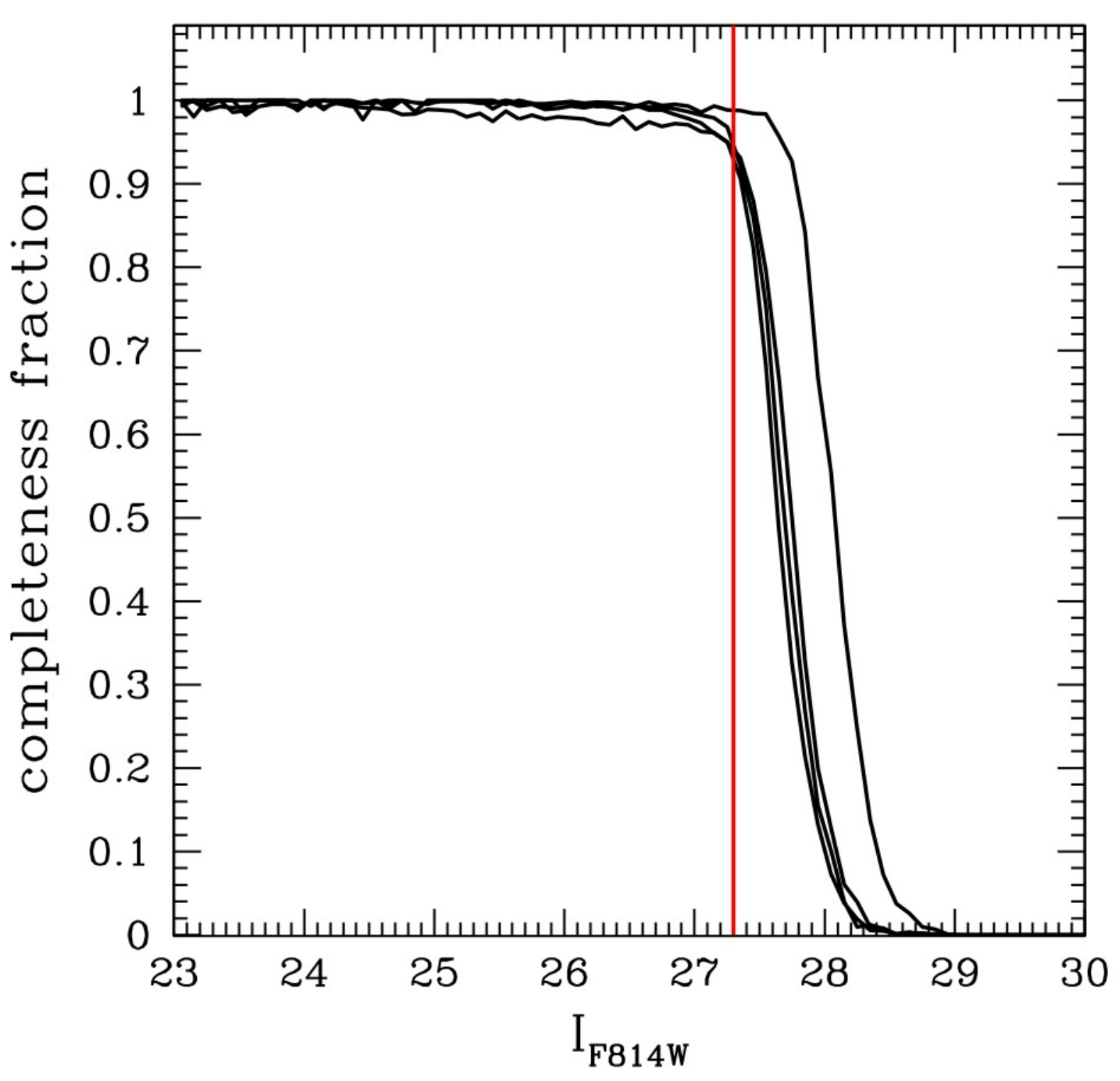}
      \caption{Completeness of point source detection as a function of $I_{F814W}$ magnitude for the four control fields.  The red line shows the  expected turnover magnitude of the globular cluster luminosity function at $I _{F814W} \simeq 27.3$ mag.  Detection of point sources is nearly 100\% complete down to this magnitude in all four control fields.  The completeness fraction at faint magnitudes is greatest for the GO-9411 field, which has the longest total integration time.
              }
         \label{controlcompleteness}
   \end{figure}
%



In Figure \ref{LF}  we plot the number of detected point sources as a function of $I_{F814W}$ magnitude for all fields.
 The agreement between the four control fields is remarkably good, which suggests that they provide a reliable estimate of the number of contaminating foreground and background point sources along the line of sight to A1185.  Furthermore, the fact that the counts in the second (outer) A1185 field are statistically indistinguishable from those in the three other control fields indicates that globular clusters are not present in significant numbers in this region, and hence if an IGC population is present in A1185 then it must be concentrated towards the innermost regions. 

Averaging results from the four control fields, we conclude that the number of contaminating foreground stars and background galaxies  is $\simeq 142 \pm 12$ down to a magnitude 
$I_{F814W} = 27.3$ mag.  The number of objects detected in the A1185 core field is an order of magnitude greater than this, a total of $\sim 1300$ globular cluster candidates. 
Likewise, Figure \ref{cmd} shows the color-magnitude relation for point sources in the A1185 core field compared with that of similar objects in the outer field of A1185.  
If we consider only those objects with colors in the range  
$0.7 \leq V-I \leq 1.5$ expected for globular clusters, then there are 977 objects in the central field compared to 32 in the outer field, strong evidence of the presence of a large population of globular clusters in the core of A1185. 
 Based on  our detection of 1422 globular cluster candidates brighter than  $I_{F814W} = 27.3$ mag, after subtracting the 
 expected number of contaminating background objects and doubling the resulting number to include the faint half
 of the GCLF, we estimate that $\sim 2560$ globular clusters in total are present in the {\em ACS} field.

   \begin{figure}
   \centering
   \includegraphics[width=9cm]{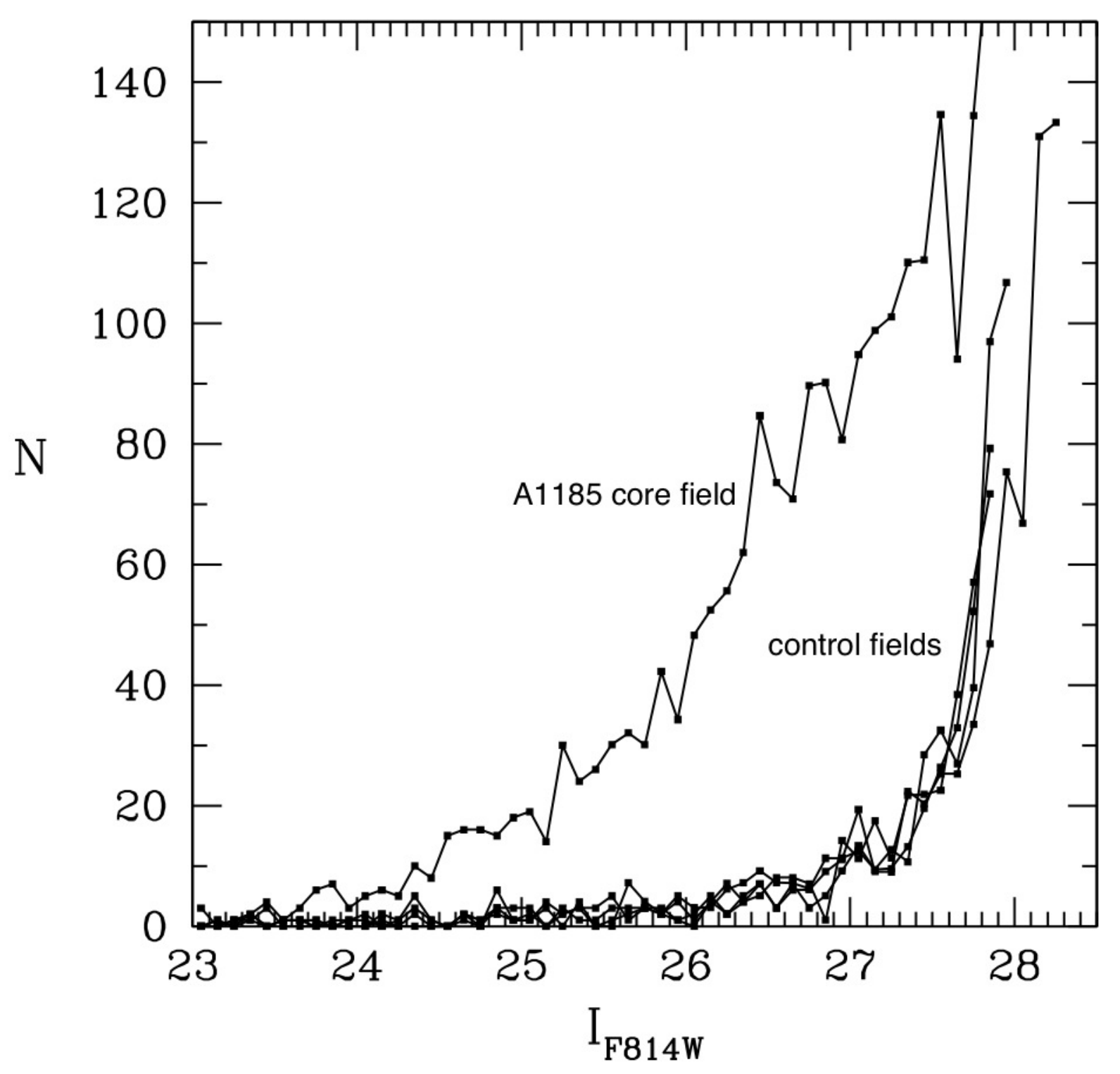}
      \caption{The number of detected point sources in the A1185 core field and the four control fields.  Counts have been corrected for incompleteness at each magnitude.    A large excess of objects is clearly seen at all magnitudes in the core field compared to the control fields, consistent with the presence of a substantial population of globular clusters.
              }
         \label{LF}
   \end{figure}
%

   \begin{figure}
   \centering
   \includegraphics[width=9cm]{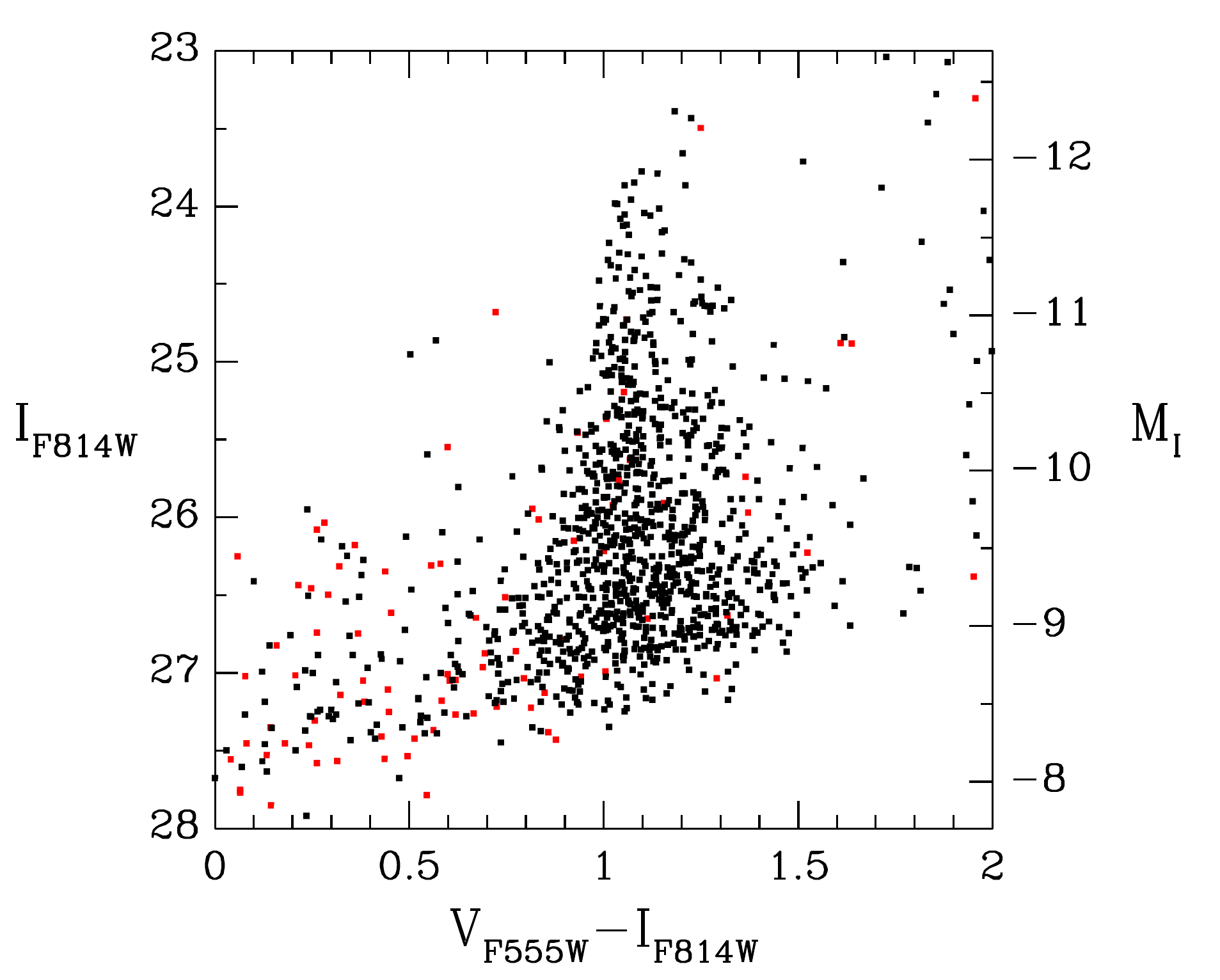}
      \caption{Color-magnitude diagram for globular-cluster-like objects in A1185 core field.  Black points are objects in the core field and red points are from the outer field indicated in Fig. \ref{CFHT}.  The prominent swath of objects with colors $0.7 \le V_{F555W}-I_{F814W} \le 1.5$ shows the detection of a substantial number of globular clusters in the core field.   Within this color range there are 
      977 objects in the central field compared to only 32 in the outer field, a factor of more than 30 enhancement in the number of objects. $M_I$ denotes the absolute magnitude of detected objects assuming a distance modulus $m-M = 35.7$.
              }
         \label{cmd}
   \end{figure}
%

To estimate the fraction of the detected globular clusters that are associated with galaxies, we performed detailed simulations of the expected distribution of globular clusters belonging to all galaxies brighter than $V \simeq 18$ ($M_V \simeq -18$) within $3\arcmin$ of the centre of the A1185 core field (fainter or more remote galaxies are unlikely to contribute contaminating globular clusters in significant numbers).  These 14 galaxies and their properties are listed in Table 1 
along with their apparent $V$-band magnitudes measured from the CFH12k images of Andreon et al. \cite{andreon2007} using SExtractor and their redshifts when known.
Although two of these galaxies do not have measured redshifts, we have assumed that all are members of A1185.  The following procedures and input parameters were used to generate the simulations:

\begin{itemize}
\item
It is well established that the number of globular clusters belonging to a galaxy scales with its luminosity.  A standard measure is the specific globular cluster frequency, $S_N$, defined as 
$$ S_N = N_{gc} \, 10^{-0.4 (M_V + 15)} \eqno(2)$$
where $N_{gc}$ is the total number of globular clusters and $M_V$ is the galaxy's absolute V-band magnitude (Harris \& van den Bergh \cite{harris1981}).   Blakeslee, Tonry \& Metzger 
\cite{blakeslee1997} measured the specific frequency of the dominant elliptical, NGC 3550, to be $S_N \simeq 6.5$, which is in the normal range for a galaxy of its luminosity.  For the other galaxies, none of whose globular cluster systems have been studied to date, we used the empirical relation between specific frequency and V-band galaxy luminosity determined by Peng et al. \cite{peng2008} from observations of 100 early-type galaxies in the ACS Virgo Cluster Survey (C\^ot\'e et. al \cite{cote2004}), with $S_N$ values ranging from $\simeq 1-5$ and the V-band magnitudes measured directly from the CFH12k images of Andreon et al. \cite{andreon2007} using SExtractor.  
This relation agrees well with that found for other galaxies (see, e.g., Ashman \& Zepf \cite{az1998} and references therein) and hence provides a reliable input for the simulations.
A total number of globular clusters was assigned to each galaxy in Table 1 based on its luminosity and assumed specific frequency.  

\medskip
\item
Each simulated globular cluster was assigned a luminosity by randomly sampling a Gaussian luminosity function  with a mean absolute magnitude of $M_I  = -8.4$ mag and a standard deviation $\sigma = 1.4$ mag, which is also consistent with previous observations of the GC population of NGC 3550 (Blakeslee et al. \cite{blakeslee1997}).   Clusters whose apparent magnitude at the distance of A1185 would be fainter than our $I_{F814W} = 27.3$ mag completeness limit were then removed from further consideration.  
Although Jord\'an et al. \cite{jordan2007} and Villegas et al. \cite{villegas2010} have shown that the properties of the GCLF correlate with parent galaxy luminosity, it is nearly uniform among high-mass galaxies that contribute the bulk of contaminating globular clusters, with turnover magnitudes and dispersions in agreement with the values adopted here.  The five most luminous galaxies in Table 1, for example, which contribute the lion's share of halo globular clusters, would be expected to have $\sigma$ ranging from $\sim 1.2$ to 1.4 mag.   Hence, any non-universality of the GCLF can safely be ignored in the present analysis.
\medskip
\item
The spatial distribution of globular clusters in galaxy halos varies widely but is generally more extended than that of the starlight (e.g., Harris \cite{harris1991}; Ashman \& Zepf \cite{az1998}; Brodie \& Strader \cite{bs2006}).  We assumed that the globular cluster system around each galaxy has a spherical distribution that follows a S\'ersic (\cite{sersic1968}) profile
$$I(r) = I_e\, {\rm exp} [-b_n(r/r_e)^{1/n}-1] \eqno(3)$$
where $I_e$ is the projected surface brightness at the effective radius, $r_e$, that contains half the total light of the system, $n$ is the S\'ersic index that determines the overall shape of the profile, and $b_n$ is a parameter related to $n$ (see Graham \& Driver \cite{graham2005} and references therein).  
Surface brightness profiles for the stellar distribution in each galaxy were measured on the CFH12k R-band image of A1185 (Fig. \ref{CFHT}) using the {\tt ellipse} routine in IRAF, and S\'ersic parameters were then obtained by 
fitting the profiles to equation 3 using $\chi^2$ minimization.\footnote{As Fig. \ref{ngc3550} shows, NGC 3550 is actually a merging system of three galaxies within a common envelope, a brightest cluster galaxy caught in the act of formation.   The light from the two companion galaxies was subtracted before measuring the underlying surface brightness profile of NGC 3550.}
As an initial guess, the simulations assumed that the effective radius of each galaxy's globular cluster distribution is twice that of its stellar distribution (Peng et al. 2011, in preparation).

\medskip
Tidal forces from the gravitational field of A1185 and galaxy-galaxy interactions are likely to limit the radial extent of each galaxy's globular cluster distribution (e.g., Merritt \cite{merritt1984}, although see 
Mihos et al. \cite{mihos2005} and Janowiecki et al. \cite{janowiecki2010}).
For this reason we simulated three different scenarios.  For one, the globular cluster distribution was not truncated, which likely leads to an {\em overestimate} of the amount of contamination from halo globular clusters in our ACS field.  For the other cases, the globular cluster population of each galaxy was truncated beyond a radius of 50 kpc or 100 kpc.
Previous observations by Blakeslee (\cite{blakeslee1996}) showed that 
the number
density of point sources surrounding NGC 3550 becomes constant beyond a projected distance of $100\arcsec$, suggesting that this galaxy's globular cluster population does not extend beyond $\sim 63$ kpc.  Similarly, in a wide-field imaging survey Rhode \& Zepf (\cite{rhode2001}; \cite{rhode2004}) found that the maximum radial extent of the globular cluster 
systems of four early-type galaxies in the Virgo cluster ranged from $\sim 30-100$ kpc.
The assumed radial distribution of globular clusters in galaxy halos is the greatest uncertainty in these models.
\end{itemize}

\begin{table*}
\caption{Galaxies near the center of A1185 used to estimate contamination in the {\em ACS} field from halo globular clusters}             
\label{table:1}      
\centering                          
\begin{tabular}{l c c c c c c c}        
\hline\hline                 
Galaxy & RA (2000) & DEC (2000) & V (mag)  & $z$ \\    
\hline                        
NGC 3550 & 11:10:38.7 & 28:46:06 & 13.2 & 0.0346  \\      
NGC 3552 & 11:10:42.8 & 28:41:35 & 14.3 & 0.0331 \\      
CGCG 155-081 & 11:10:37.5 & 28:43:58 & 14.9 & 0.0310  \\      
MCG+05-27-004 & 11:10:52.0 & 28:42:09 & 15.3 & 0.0340   \\ 
NGC 3553 & 11:10:40.5 & 28:41:05 & 15.3 & 0.0326   \\      
2MASX J11105941+2844424 & 11:10:59.4 & 28:44:43 & 15.7 & 0.0327  \\    
2MASX J11104649+2841338 & 11:10:46.5 & 28:41:34 & 15.9 & 0.0326  \\      
2MASX J11103987+2842248 & 11:10:39.9 & 28:42:25 & 16.3 & 0.0288   \\      
2MASX J11103964+2845208 & 11:10:39.6 & 28:45:21 & 16.9 & 0.0383  \\      
2MASX J11103683+2842138 & 11:10:36.8 & 28:42:14 & 16.9 & $-$   \\      
2MASX J11104185+2845408 & 11:10:41.9 & 28:45:41 & 17.0 & $-$   \\      
2MASX J11105850+2842384 & 11:10:58.5 & 28:42:39 & 17.1 & 0.0346   \\      
SDSS J111035.04+284205.2 & 11:10:35.0 & 28:42:05 & 17.3 & 0.0338  \\      
2MASX J11103767+2841528 & 11:10:37.7 & 28:41:53 & 17.6 & 0.0343   \\      
\hline                                   
\end{tabular}
\end{table*}

\noindent Using the procedures and input parameters described above, 100 random realizations of the expected distribution of globular clusters in the halos of the galaxies in Table 1 were generated and the simulated clusters brighter than $I_{F814W} \leq 27.3$ mag that fall within the {\em ACS} field boundaries were counted.  An example is shown in Fig. \ref{models}.  

Based on these simulations, we estimate that $\sim711 \pm 26$ IGC candidates in the A1185 core field are likely to be globular clusters in galaxy halos if their spatial extent in unlimited, most of them belonging to NGC 3550.   We emphasize, however, that this probably overestimates the number of contaminating halo globular clusters.  For the case in which each galaxy's globular cluster system is assumed to be tidally truncated beyond 100 kpc, the estimated number of contaminating objects is $\sim 472 \pm 30$, and for a limiting radius of 50 kpc  the number falls to $\sim 166 \pm 14$.  

Combining the estimated contamination from globular clusters in galaxy halos and the expected number of unassociated objects along the line of sight, we estimate that less than half of the 1422 IGC candidates brighter than $I_{F814W} \simeq 27.3$ mag can be accounted for this way, and hence the majority of detected objects are likely to be genuine IGCs.   Our best estimate, based on the assumption that globular clusters are not found beyond 100 kpc in galaxy halos, is that  $\sim 1616 \pm 91$ IGCs of all magnitudes exist in the core {\em ACS} field.\footnote{This estimate of $\sim 1616$ IGCs is based on the  
detection of 1422 objects down to $I_{F814W} \simeq 27.3$ mag, subtracting 142 background objects brighter than 
this same limiting magnitude and 
assuming that 472 of the IGC candidates are globular clusters in galaxy halos.  The resulting number is then 
doubled  to account for the fact that we observe only the brighter half of the GCLF.  Formal statistical uncertainties in the number of detected IGCs, unrelated point sources, and halo globular clusters have been added in quadrature, 
however systematic errors resulting from our lack of knowledge about the globular cluster 
systems of individual galaxies make the results considerably more uncertain. }

However, although there is no doubt that we have detected a large population of globular clusters in the core of A1185, it is difficult to say 
with certainty whether any of them are truly intergalactic. 
In addition to the aforementioned assumptions and uncertainties inherent in our simulations, other factors might also plausibly raise the number of globular clusters associated with galaxies.  For example,  although  low- or intermediate-luminosity galaxies (fainter than  $M_V \simeq -18$) have not 
been included in the simulations because their globular cluster populations are expected to be sparse, a large population of such galaxies could collectively contribute several hundred additional contaminating halo clusters
scattered over the {\em ACS} field.  Uncertainties in the magnitudes of the galaxies in Table 1 could also raise or lower the expected number of galactic globular clusters.

It is, of course, possible to boost the predicted number of contaminating objects to match the number of detected objects in the {\em ACS} field by arbitrarily increasing the globular cluster populations of the galaxies or changing their radial distributions.  For example, doubling the assumed specific 
frequency of every galaxy in Table 1 and assuming that assuming that the globular cluster
distribution is twice as extended as the galaxy light yields an estimated $1292 \pm 34$ contaminating halo globular clusters in the {\em ACS} field.  However, such a model seems unlikely, especially in light of the measured specific frequency of NGC 3550 (Blakeslee et al. \cite{blakeslee1997}).  
To further assess the validity of such a high specific frequency model, the simulated distributions of globular clusters (including the addition of $142 \pm 12$ randomly distributed foreground/background objects) were compared to  the observed distribution using a two-dimensional Kolmogorov-Smirnov test (Peacock \cite{peacock1983}; Fasano \& Franceschini \cite{fasano1987}) as implemented by Press et al (\cite{press1992}).  Based on this analysis, the probability that the simulated and observed globular cluster distributions could both be drawn from the same parent distribution is much less than 1\%, although one must bear in mind the limitations of these models.

   \begin{figure}
  \centering
   \includegraphics[width=9cm]{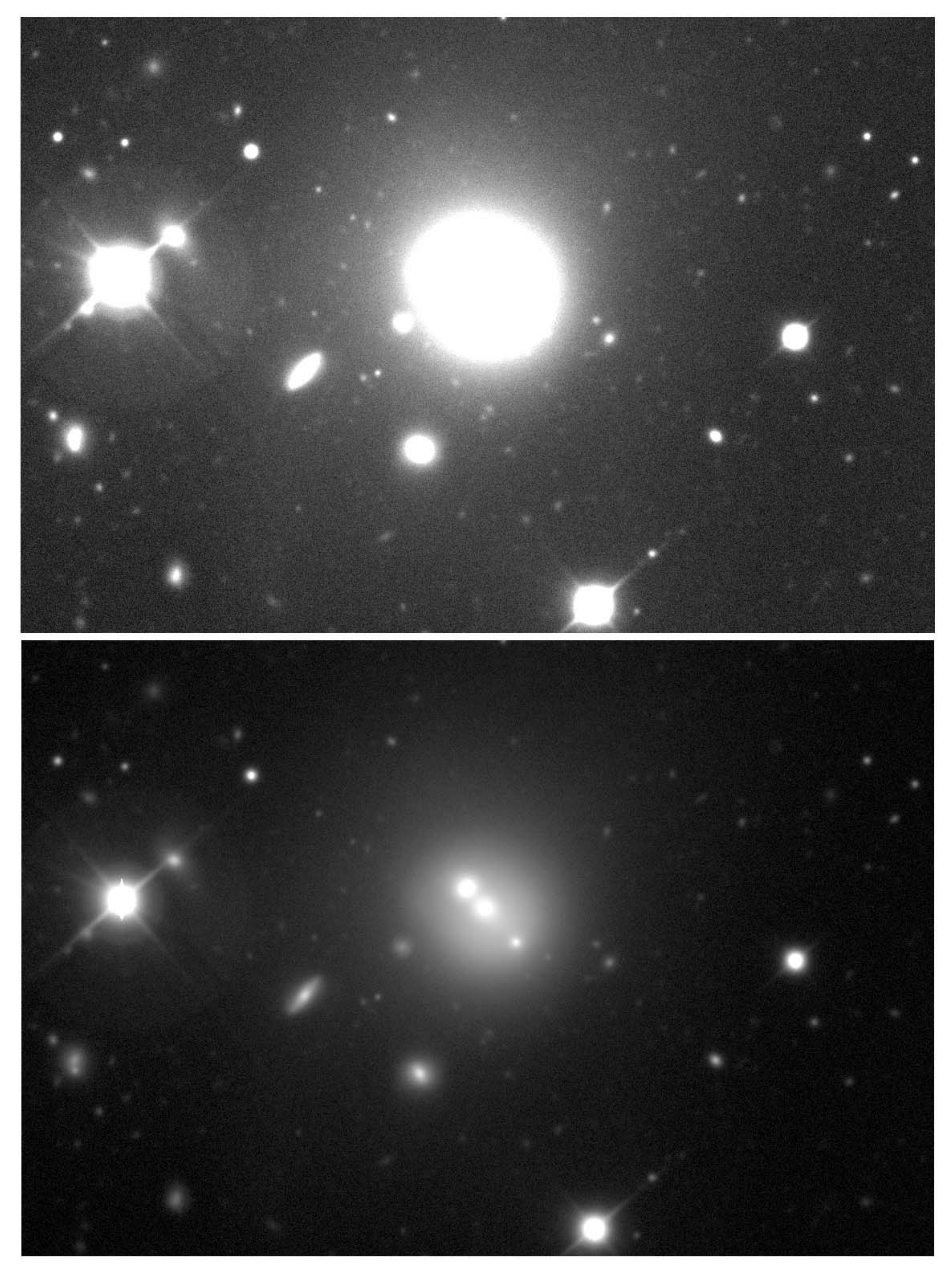}
     \caption{NGC 3550, the brightest member galaxy in A1185.  The image on top is  
      displayed with a linear intensity scale and NGC 3550 appears as a single large elliptical galaxy.  The logarithmic display below  
      reveals that NGC 3550 is actually a complex system of three smaller galaxies within a common envelope.
              }
        \label{ngc3550}
   \end{figure}

   \begin{figure*}
   \centering
   \includegraphics[width=18.5cm]{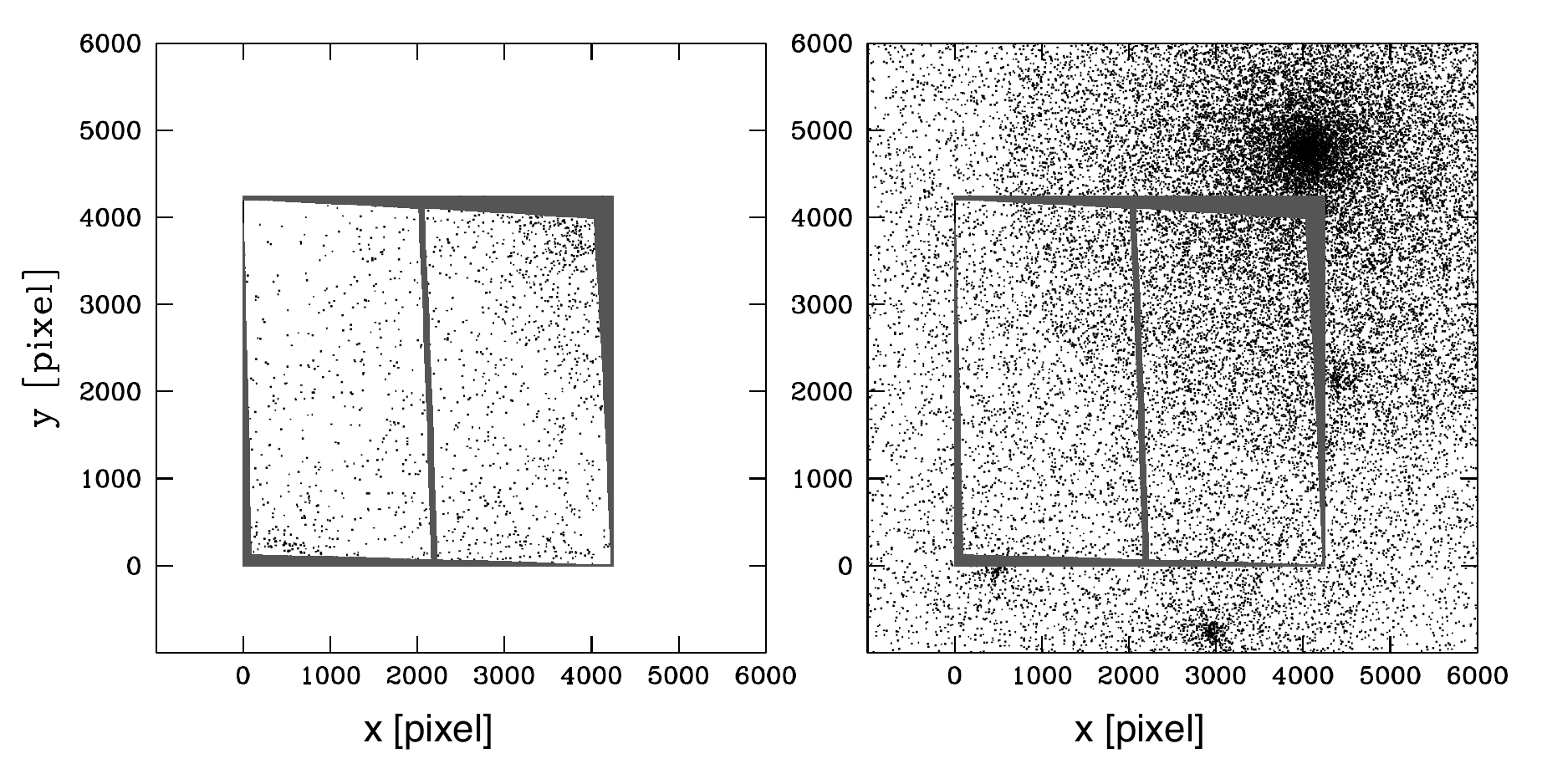} 
     \caption{Simulations of the expected contamination from globular clusters in halos of galaxies that lie near the A1185 core field.  The left panels shows the real distribution of globular clusters brighter than $I_{F814W} = 27.3$ mag in the {\em ACS} image of the heart of 1185.  
The right panel shows a superposition of 10 simulated distributions of halo globular clusters  associated with the galaxies listed in Table 1 down to the same $I_{F814W} = 27.3$ mag limit and assuming {\em no} truncation of their radial distribution, as described in the text.  
 }
     \label{models}%
    \end{figure*}
%

   \begin{figure}
  \centering
   \includegraphics[width=9.0cm]{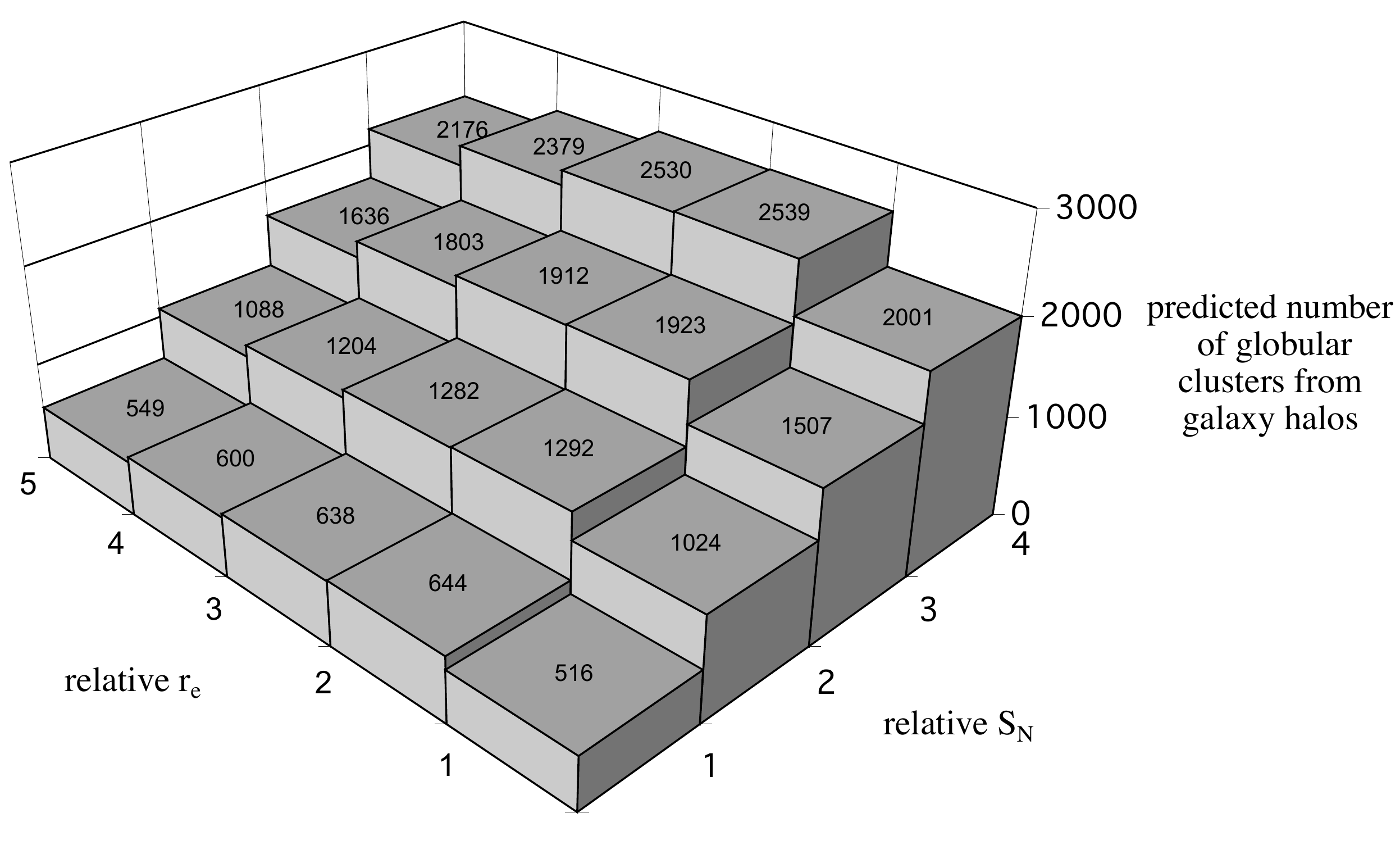}
   \caption{An illustration of the expected contamination from globular clusters in galaxy halos near the center of A1185 and its dependence 
   on the assumed properties of those systems.  
   The globular cluster population of each galaxy in Table 1 was simulated by  assuming that its radial distribution is described by the same Sersic index $n$ as the galaxy's stellar light, but with an effective radius that is some 
   multiple of the galaxy's (the axis labelled `relative $r_e$').   Similarly, the number of globular clusters belonging to each galaxy was allowed to vary relative 
   to the expected number if the galaxy had a normal specific frequency for its luminosity as summarized in Peng et al. \cite{peng2008} (the axis labelled `relative $S_N$').  
   Hence, for example, a relative $r_e = 2$ means that the effective radius of a galaxy's globular cluster 
   distribution was assumed to be twice that of its stellar distribution, and an `effective $S_N = 3$' means that a galaxy was assigned three times the `normal' number of globular clusters for a galaxy of its luminosity.  
   Additional details of the simulations are discussed in the text.  The third axis shows the 
      average number of contaminating globular clusters brighter than  
   $I_{F814W} \leq 27.3$ mag predicted to fall within the central {\em ACS} field in A1185 based on many simulations.   Although the relative $r_e$ and $S_N$ were allowed to vary independently for each galaxy, for clarity the results shown in 
   this figure assume constant values of relative $r_e$ and relative $S_N$ for all galaxies. 
   These simulations 
   demonstrate that the large number of globular clusters seen in our {\em ACS} field cannot all reside in galaxy halos unless most or all of the galaxies in Table 1 have unusually rich globular cluster systems. }
     \label{stats}
    \end{figure}

Given the uncertainties in the simulations, and to explore the full range of parameter space $-$ including more extreme possibilities $-$ several thousand additional simulations were run in which the parameters of each galaxy's globular cluster system were randomly chosen in the range $1 \leq S_{N} \leq 20$, $0^{\prime\prime} \leq r_e \leq 200^{\prime\prime}$ and $0 \leq n \leq 10$.  
These experiments  demonstrate that, even in the most favorable case in which the spatial extent of each galaxy's GC system is not tidally limited,   
 a minimum necessary condition for  
$\sim 1000-1500$ contaminating halo globular clusters to fall within our central ACS field is 
 that two or more of the brightest member galaxies (NGC 3550, NGC 3552, CGCG 155-081 and MCG +05-27-004) would have to possess extremely rich globular cluster systems, with $S_N \geq 10-20$ (see Fig. \ref{stats} for an 
 illustration).   While this possibility cannot be ruled out, it seems unlikely that A1185 would 
 have the distinction of being the only known cluster with multiple high $S_N$ galaxies, especially given the unexceptional size of NGC 3550's globular cluster population (Blakeslee et al. \cite{blakeslee1997}).

We conclude that, for any reasonable assumptions, it is difficult to account for the large number of globular clusters detected in our central {\em ACS} field without including a significant population of IGCs that do not belong to galaxies.  
Although we cannot completely rule out the possibility that most or all of the globular clusters detected in the core of A1185 could be associated with galaxies, this seems unlikely.  
Further insights into the nature of these objects could come from their luminosities and colors, which are examined in more detail below.

\subsection{The globular cluster luminosity function}

The luminosity function (or, equivalently, the mass function) is a fundamental property of globular cluster systems.  The present-day GCLF has been shaped by many factors, including physical conditions at the time of their formation, internal processes such as stellar mass loss and two-body relaxation, and stripping or destruction of clusters in galaxy environments as a result of disk shocking, tidal forces, and dynamical friction.  Given this variety of influences, it is remarkable that 
 the GCLF in galaxies can, to first order, be described universally by a Gaussian distribution with a maximum at $M_I \simeq -8.4$ and a dispersion of $\sigma \simeq 1.2-1.4$ magnitudes (e.g., Harris \cite{harris1991}).   
 However, some correlation is observed between GCLF properties and parent galaxy luminosity (Jord\'an et al. \cite{jordan2006}, \cite{jordan2007}; Villegas et al. \cite{villegas2010}), which suggests that environment plays a role in the formation and/or evolution of globular cluster systems.
 Several authors have shown, for example, that an initial power-law mass function can be transformed into one with a log-normal shape as a consequence of cluster disruption and evaporation in galaxies (see, e.g.,  Fall \& Zhang \cite{fall2001}; Vesperini et al.  \cite{vesperini2003}; McLaughlin \& Fall \cite{mclaughlin2008}; Elmegreen \cite{elmegreen2010}).   
 
It therefore seems plausible that the luminosity function of IGCs born outside of galaxies might differ from that of globular clusters which formed and evolved within galaxies, although it is hard to predict {\it a priori} what form such differences might take.   If, on the other hand, IGCs originated in galaxies and were subsequently stripped from halos or spilled into intergalactic space during the disruption of their parent galaxies then the IGC luminosity function is likely to be similar to that of galactic globular clusters.  Hence the luminosity function of globular 
 clusters in the core of A1185 might provide some insights into their origin.

Figure \ref{GCLF} shows the observed luminosity function of globular cluster candidates based on the deep F814W observations.  The number of objects in each luminosity bin has been corrected by subtracting the expected contamination from foreground stars and unresolved background galaxies (Fig. \ref{LF}) and correcting for incompleteness (Fig. \ref{completeness}).  
Using $\chi^2$ minimization to fit the observed GCLF to a Gaussian distribution with the turnover magnitude fixed at $I_{F814W} = 27.3$ mag but with the amplitude and width of the distribution as free parameters yields a best fit with $\sigma \simeq 1.2 \pm 0.2$, which is shown for comparison in the figure.  Within the uncertainties, this falls well  
within the normal range for GCLFs in galaxies.  

   \begin{figure}
  \centering
   \includegraphics[width=9.0cm]{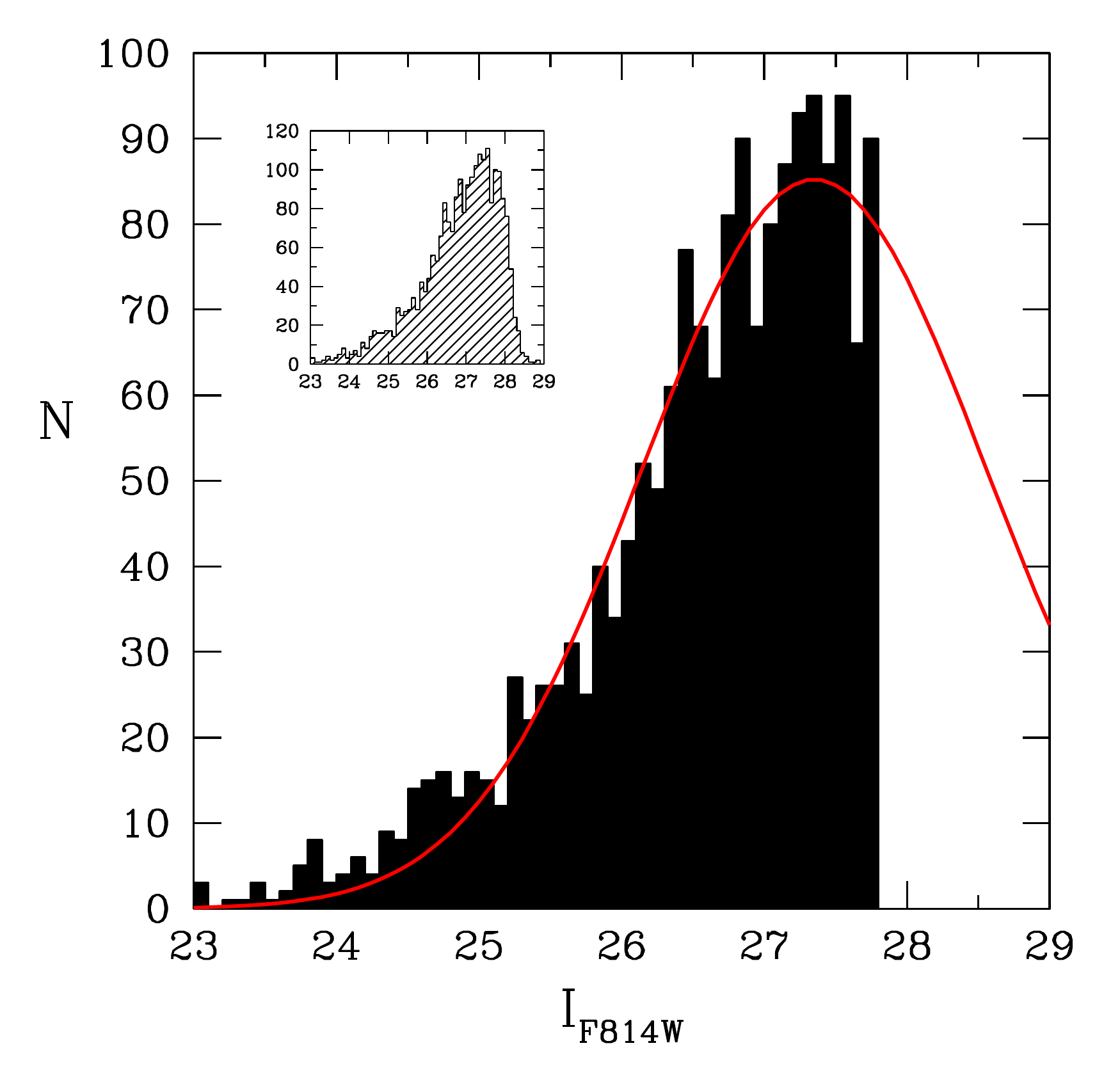}
     \caption{The observed luminosity function of globular clusters in the core of A1185.  The data have been corrected for incompleteness and the expected number of contaminating foreground/background objects have been removed based on the averaged counts at each magnitude from the four control fields.  The luminosity function data are shown only to $I_{F814W} \simeq 27.8$ mag because beyond that incompletenss corrections become too large (more than a factor of two) to be reliable.  The red curve is the best-fitting Gaussian luminosity function with a peak luminosity at $I_{F814W} = 27.3$ mag and a standard deviation $\sigma = 1.2$, as discussed in the text.  The observed luminosity function is clearly consistent with the standard GCLF observed in many galaxies.  The inset shows the raw luminosity function of 
     globular cluster candidates without any corrections for background contamination or incompleteness.
                   }
        \label{GCLF}
   \end{figure}

\subsection{The globular cluster color distribution}

It is well established that the colors of old globular clusters, which are determined primarily by their metallicities, show a dependence on parent galaxy luminosity, in the sense that luminous galaxies have redder (more metal-rich) globular clusters on average than fainter galaxies (e.g., van den Bergh \cite{vdb1975}; Brodie \& Huchra \cite{brodie1991}; Forbes \& Forte \cite{forbes2001}; C\^ot\'e, Marzke \& West \cite{cmw1998}; Peng et al. \cite{peng2006}).   
Hence the color distribution of globular clusters in the core of A1185 could, in principle, provide some constraints on the numbers and types of donor galaxies from which an IGC population might have been stripped (e.g., Forbes, Brodie \& Grillmair \cite{forbes1997}; C\^ot\'e, Marzke \& West \cite{cmw1998}; Larsen et al. 2001; Gebhardt \& Kissler-Patig \cite{gebhardt1999}; Coenda et al. \cite{coenda2009}).  
In reality, however, such constraints are likely to be weak because of degenerate dependencies of globular cluster colors on galactocentric distances and host galaxy properties.
A preponderance of blue IGCs could arise, for example, if the majority came from dwarf galaxies that were tidally disrupted {\it or} if they were stripped from the outer halos of giant galaxies whose globular clusters are bluer on average than their counterparts closer to the galaxy center (Geisler et al. \cite{geisler1996}; Rhode \& Zepf \cite{rhode2004}; Bassino et al. \cite{bassino2006}; Tamura et al. \cite{tamura2006}).   Nevertheless, if the globular clusters are found to have a distinctive color distribution then this could be an important clue to their origin.
  
Artificial star experiments show that matched detection of objects added to both the F555W and 
F814W images is 100\% complete to $I_{F814W} \simeq 26.5$ mag, as the shallower F555W observations result in fewer detections at fainter magnitudes.  To minimize the effect of this incompleteness on the inferred color distribution of globular clusters, we consider only those objects brighter than this magnitude limit.  

Figure \ref{colors} shows the observed color distribution of globular clusters in the A1185 core field.   These colors are quite similar to those of globular clusters associated with galaxies (Gebhardt \& Kissler-Patig \cite{gebhardt1999}; Peng et al \cite{peng2006}; Forbes et al. \cite{forbes2007}), skewed strongly toward bluer (more metal-poor) colors.  
The peak in the color distribution occurs at $V-I \simeq 1.05$, in excellent agreement with the results of 
Gebhardt \& Kissler-Patig \cite{gebhardt1999}, who found peak colors around $V-I \simeq 0.94-1.13$ depending on host galaxy luminosity.
Although it is impossible to distinguish between IGCs and halo globular clusters in the {\em HST} images, it is clear that if IGCs are present then their colors cannot be grossly different from those of normal globular clusters.  The colors are predominantly blue, as expected if IGCs are the surviving orphans of disrupted dwarf galaxies or were stripped from the outer halos of larger galaxies.   

Figure \ref{colors} hints at possible bimodality or asymmetry of the globular cluster color distribution.  To quantify this impression, we used the KMM mixture mode algorithm (McLachlan \& Basford \cite{mclachlan1988}) as implemented by Ashman, Bird \& Zepf \cite{abz1994}.  KMM uses a
maximum likelihood estimator to determine the best fitting mixture of one or more Gaussian components to an observed 
distribution and estimates the mean and fraction of data points belonging to each component.
Because outliers can adversely affect the KMM results (see Ashman et al. \cite{abz1994} for a discussion) we restricted our analysis to the 616 clusters with colors in the range $0.8 \leq V-I \leq 1.4$ and $I_{F814W} \le 26.5$ mag.
Assuming homoscedasticity (i.e., the same variance for each subpopulation), the KMM algorithm strongly rejects (at a confidence level of more than $99\%$) the hypothesis that the color distribution seen in Fig. \ref{colors} is adequately described by a single Gaussian.
The best fitting model has two Gaussian components, with roughly 2/3 of globular clusters belonging to a blue subpopulation 
with a mean color $\langle V-I \rangle \simeq 1.04$ and 1/3 belonging to a redder subpopulation with  $\langle V-I \rangle \simeq 1.23$.    

Given the detection of distinct blue and red subpopulations, we looked for evidence of the so-called `blue tilt'  in the color-magnitude relation (Fig. \ref{cmd}), an observed tendency for blue globular clusters to become progressively redder at brighter magnitudes (e.g., Harris et al. \cite{harris2006}; Strader et al. \cite{strader2006}; Peng et al. \cite{peng2009}; Blakeslee, Cantiello \& Peng \cite{blakeslee2010}; Mieske et al. \cite{mieske2010}).  In principle, the strength of the blue tilt, which is usually interpreted as a mass-metallicity relation, could provide clues to the origin of the globular clusters in the center of A1185.  As noted by Mieske et al. (\cite{mieske2010}), for example, the slope of the blue tilt in the Virgo and Fornax 
clusters of galaxies varies as a function of host galaxy luminosity and galactocentric distance.  

For this purpose, we divided the sample of 616 globular clusters satisfying the aforementioned color and magnitude criteria into four luminosity bins containing equal numbers of objects.  The KMM algorithm was then used to find the
best fitting blue and red peaks in the color distribution for each luminosity bin (again assuming homoscedasticity) with $V-I = 1.04$ and 1.23 chosen 
as the initial guess for their locations.   Within each luminosity bin, the color distribution is well fitted by a mixture of two Gaussians.   Because the KMM analysis does not provide a direct estimate of the uncertainties in the derived quantities,
bootstrap resampling was used to estimate the variance in the mean colors of the blue and red subpopulations. 
Results are shown in Fig. \ref{bluetilt}.  
Despite the relatively weak sensitivity of $V-I$ colors to 
metallicity (compared to the longer wavelength baselines provided by $g-z$ or $B-I$ used in most previous studies) a blue tilt is apparent.  No obvious trend is seen for the red clusters.  These visual impressions are confirmed 
by statistical analysis, which yields a Pearson correlation coefficient $r = -0.99$ (highly significant) for the color-magnitude relation 
of the blue subpopulation but only $r = -0.45$ (not significant) for the red subpopulation.
The slope of the best fitting linear regression to the blue tilt seen 
 in Fig. \ref{bluetilt} is $d(V-I)/dI \simeq -0.039 \pm 0.013$, with the uncertainty estimated from 
 bootstrap resampling.   This slope is within the range found in previous studies.  Given 
the sizable uncertainties, however, we 
refrain from further analysis or interpretation of these results.
 
   \begin{figure}
  \centering
   \includegraphics[width=9cm]{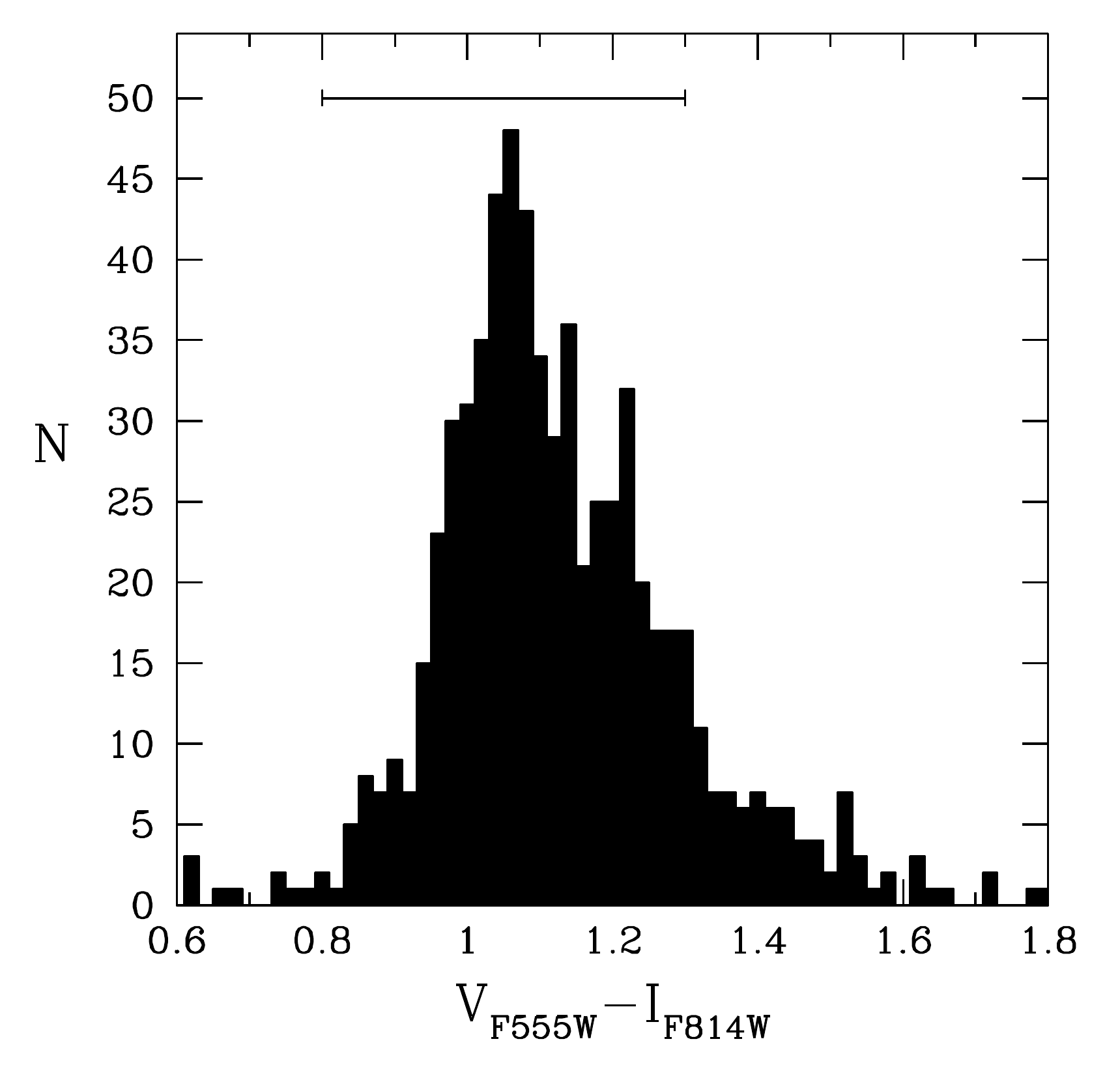}
     \caption{The observed $V-I$ color distribution of globular clusters in the core of A1185. The bar indicates typical
     colors of globular clusters in the sample of Gebhardt \& Kissler-Patig (\cite{gebhardt1999}), corresponding 
     to the FWHM of their color distribution).
              }
        \label{colors}
   \end{figure}

   \begin{figure}
  \centering
   \includegraphics[width=9cm]{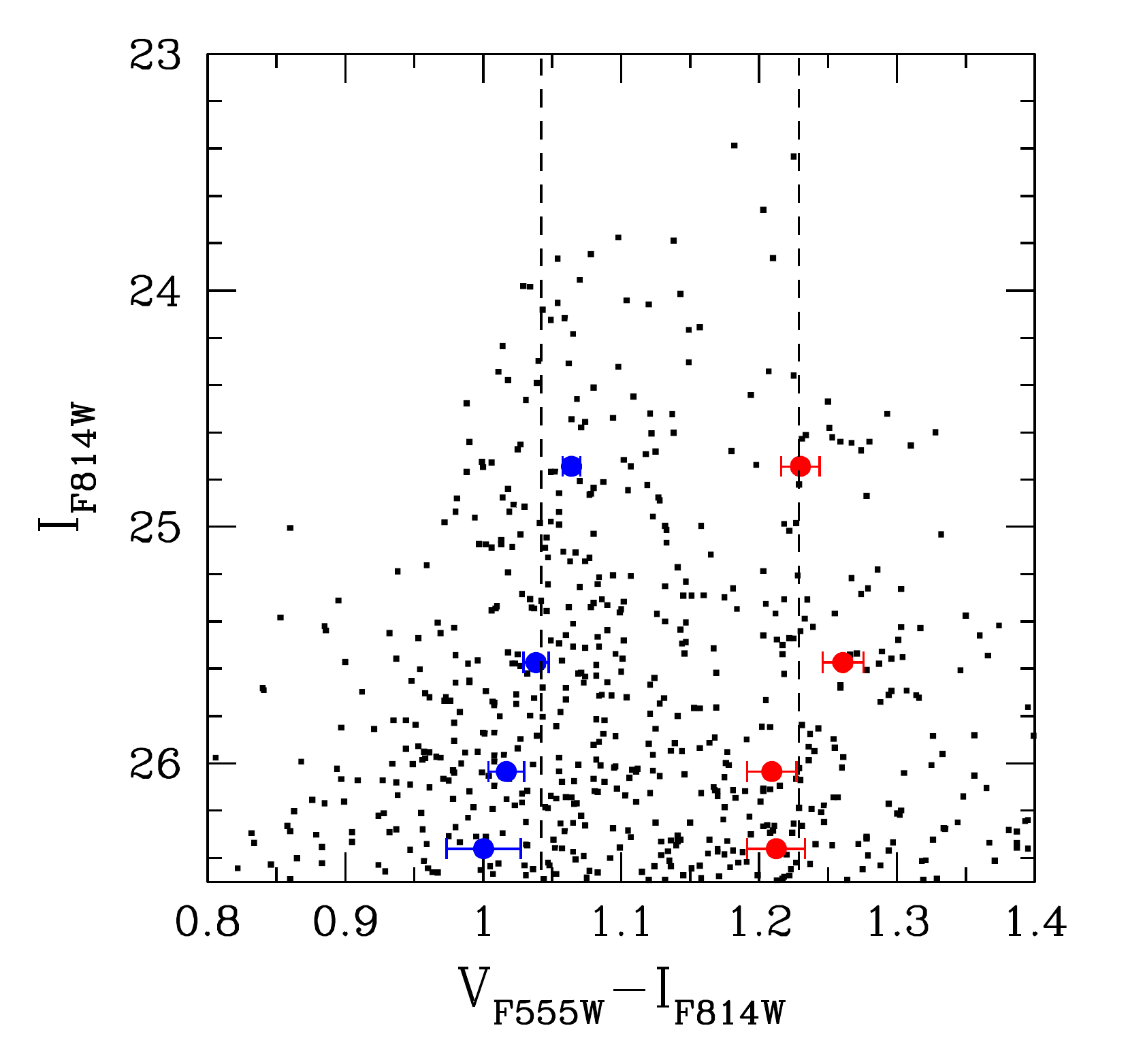}
     \caption{Color-magnitude diagram for 616 globular clusters in the core of A1185 within the range $0.8 \leq V-I \leq 1.4$ and $I_{F814W} \le 26.5$ mag.   The blue and red points indicate the locations of the blue and red 
     peaks determined by KMM fitting of the color distribution in four different luminosity bins to a model with two Gaussian components, as described in the text.  Uncertainties in the peak locations were calculated 
     via bootstrap resampling.  The dashed lines denote the locations of the blue and red 
     peaks found from KMM analysis for the entire sample of 616 clusters.  A correlation between color and 
     magnitude is apparent for the blue globular clusters (dubbed the 'blue tilt') while no such trend is seen for the red clusters.
              }
        \label{bluetilt}
   \end{figure}

\section{Discussion}

The results presented here show that the $\sim 1300$ globular clusters found in our {\em ACS} field are indistinguishable, at least in terms of their optical colors and luminosities, from the metal-poor globular cluster populations observed in hundreds of galaxies to date.  
Although a non-negligible fraction of these objects are undoubtedly chance projections of globular clusters residing in the extended halos of galaxies, we have argued based on detailed simulations that the majority are 
likely to be intergalactic.  Nevertheless, given the uncertainties in these simulations we cannot confidently rule out the possibility that most or even all of the detected globular clusters might belong to 
galaxies rather than roaming the space between them.

Confirmation that the candidate IGCs are truly intergalactic could be obtained in several ways.  
The most direct, but observationally challenging, method would be to measure velocities of these objects to determine 
whether they are indeed gravitationally bound to individual galaxies or whether they constitute 
an independent dynamical system in the core of A1185. In particular, the largest potential source of contaminating halo 
globulars, NGC 3550, has an observed radial velocity of 10,520 km s$^{-1}$ (Tonry \cite{tonry1985} gives velocities of 10447, 10388 and 1100 km s$^{-1}$ for each of the three components seen in Fig.  9), which differs by $\sim 1200$ km s$^{-1}$ from the 9300 km s$^{-1}$ mean velocity of the A1185 system (Mahdavi et al. \cite{mahdavi1996}), 
consistent with the spatial and kinematic substructure seen in this dynamically young cluster.   Consequently, halo globular clusters associated with NGC 3550 could be readily distinguished from 
genuine IGCs moving in the gravitational potential of A1185 on the basis of their expected large velocity 
differences. 
The faint magnitudes of even the brightest IGC candidates in A1185 make such observations challenging, though not impossible, with 8-m to 10-m class telescopes.

Deep images of additional fields surrounding NGC 3550 could also help to determine whether IGCs are present.  Globular clusters that are gravitationally bound to galaxies should generally be distributed symmetrically about their centers.
If the number density of globular clusters in the central A1185 field shown in Fig. \ref{CFHT} is found to be significantly higher than that in other fields equidistant from NGC 3550, then this would support the idea that they are intergalactic.  Several strategically placed fields might even allow the centroid of the putative IGC population to be localized. 

Additional clues to the nature and origin of the globular cluster population in the heart of A1185 could come from measurements of their ages and metallicities.  Although it is difficult at present to obtain such information spectroscopically because of the distance to A1185, a combination of optical and near-infrared colors can provide an alternative measure of ages and metallicities (e..g, Puzia et al. \cite{puzia2002}; Hempel et al. \cite{hempel2007}; Carter et al. \cite{carter2009}). 

Finally, we note that further study of A1185 may offer a tantalizing opportunity to  glimpse the ongoing creation of IGCs.  The merging galaxy system Arp 105 (see Fig. \ref{CFHT}) is ejecting a long, narrow plume of material that probably includes extant globular clusters destined to become future intergalactic vagabonds, as well as newly 
created dwarf galaxies and possibly young star clusters (Duc \& Mirabel \cite{duc1994}).  Similar tidal streams in other nearby galaxy clusters (e.g., Gregg \& West \cite{gregg1998}; Calc\'aneo-Rold\'an et al. \cite{calcaneo2000}; Rudick et al. \cite{rudick2009}) would be excellent places to search for other IGCs.  Deep, large-area surveys such as the ongoing {\it Next Generation Virgo Cluster Survey} (Ferrarese et al. 2010, in preparation) and the {\em ACS Treasury Survey of the Coma Cluster} (Carter et al. \cite{carter2008}) will also provide rich databases that can be searched for IGCs.  

\section{Conclusions}

Using deep {\em HST/ACS} images, we have confirmed the presence of several thousand globular clusters in the heart of the rich galaxy cluster A1185, in a field that is not centered on any bright galaxy.  Simulations suggest that the majority of these objects may be intergalactic, although the large uncertainties in these models make it impossible to reach firm conclusions about the reality or size of such a population.  Our best estimate is that a total of $\sim 1600$ IGCs reside in the core {\em ACS} field, although we cannot rule out the possibility that they could be gravitationally bound to galaxies.  The colors and luminosities of these objects are similar to those of metal-poor globular clusters found in galaxies, suggesting that $-$ if they are indeed  
intergalactic $-$ they probably formed in galaxies and were removed later by tidal stripping or other dynamical processes. 
We suggest that future observations, especially velocity measurements, could establish the true nature of the globular clusters in the center of A1185 as intergalactic wanders or 
galactic residents.  

\begin{acknowledgements}
We thank Stefano Andreon and Jean-Charles Cuillandre for kindly providing their reduced CFH12k images of A1185, Sidney van den Bergh for enlightening discussions, and the anonymous referee for suggestions that helped to improve the paper.  Support for programme HST-GO-9488 was provided by NASA through a grant from the Space Telescope Science Institute which is operated by the Association of Universities for Research in Astronomy, Inc., under NASA contract NAS 5-26555.  This research made use of the NASA/IPAC Extragalactic Database (NED) which is operated by the Jet Propulsion Laboratory, California Institute of Technology, under contract with NASA.  MJW thanks the Herzberg Institute of Astrophysics for its hospitality during much of this work and acknowledges additional support from NSF grant AST 02-05960.  AJ acknowledges support from Fondecyt project 1095213, Anillo ACT86, BASAL CATA PFB-06, FONDAP CFA 15010003 and MIDEPLAN ICM Nucleus P07-021-F.  Part of the work
reported here was done at the Institute of Geophysics and Planetary Physics, under the auspices of the 
U.S. Department of Energy by Lawrence Livermore National Laboratory in part under Contract W-7405-Eng-48 
and in part under Contract DE-AC52-07NA27344.

\end{acknowledgements}

477

\end{document}